\documentclass[a4paper,11pt]{article}
\usepackage[top=1 in,bottom=1 in,left=1.0 in,right=1.0 in]{geometry}

\usepackage{array} 
\usepackage{float}

\usepackage{amssymb,amsmath,amsthm}
\usepackage{color}
\usepackage{accents}
\usepackage{texdraw}
\usepackage{eucal}
\usepackage{todonotes}
\usepackage{epic,epsfig}
\usepackage{graphicx}
\usepackage[all]{xy}
\usepackage{tikz}

\newtheorem{remark}{Remark}
\numberwithin{equation}{section}
\DeclareMathAccent{\wtilde}{\mathord}{largesymbols}{"65}
\DeclareMathAccent{\what}{\mathord}{largesymbols}{"62}

\def\m@th{\mathsurround=0pt}
\mathchardef\bracell="0365
\def\upbrall{$\m@th\bracell$}
\def\undertilde#1{\mathop{\vtop{\ialign{##\crcr
    $\hfil\displaystyle{#1}\hfil$\crcr
     \noalign
     {\kern1.5pt\nointerlineskip}
     \upbrall\crcr\noalign{\kern1pt
   }}}}\limits}

\newcommand{\wh}{\widehat}
\newcommand{\wt}{\widetilde}

\newcommand{\bblu}{\begin{color}{blue}}
\newcommand{\bred}{\begin{color}{red}}
\newcommand{\ecl}{\end{color}}

\newcommand{\bI}{\boldsymbol{I}}

\newcommand{\bK}{\boldsymbol{K}}

\newcommand{\bM}{\boldsymbol{M}}

\newcommand{\br}{{\boldsymbol{r}}}

\newcommand{\be}{\begin{equation}}
\newcommand{\ee}{\end{equation}}
\newcommand{\bea}{\begin{eqnarray}}
\newcommand{\eea}{\end{eqnarray}}
\newcommand{\bse}{\begin{subequations}}
\newcommand{\ese}{\end{subequations}}

\newcommand{\bc}{\boldsymbol{c}}

\newcommand{\bs}{{\boldsymbol s}}

\def \h#1{\widehat{#1}}
\def \t#1{\widetilde{#1}}

\begin{document}
\title{On the Dbar method and direct linearization approach
of the lattice KdV type equations}

\author{Leilei Shi$^{1,2}$, ~~ Cheng Zhang$^{1,2}$,
	~~ Da-jun Zhang$^{1,2}$\footnote{Corresponding author.
		Email: djzhang@shu.edu.cn}
	~~\\
	{\small $~^1$Department of Mathematics, Shanghai University, Shanghai 200444,    China} \\
	{\small $^{2}$Newtouch Center for Mathematics of Shanghai University,  Shanghai 200444, China}
	}

\maketitle

\begin{abstract}

The purpose of this paper is to bridge the gap between the Dbar method and the
direct linearization approach for the lattice Korteweg-de Vries (KdV) type equations.
We develop the Dbar method  to study some discrete integrable equations in the Adler-Bobenko-Suris list.
A Dbar problem is considered to define the eigenfunctions of the Lax pair of the lattice potential KdV equation.
We show how an extra parameter is introduced in this approach
so that the lattice potential modified KdV equation and lattice Schwarzian KdV equation are derived.
We also explain how the so-called spectral Wronskians make sense in constructing
the H3$(\delta)$, Q1$(\delta)$ and Q3$(\delta)$ equations.
Explicit formulae of multi-soliton solutions are given for the derived equations,
from which one can see the connections between the direct linearization variables
($S^{(i,j)}$ and $V(p)$) and the eigenfunctions and their expansions respectively
at  infinity and a finite point.

\begin{description}
\item[Keywords:] Dbar method; direct linearization approach; lattice KdV type equation; soliton solution
\item[Mathematics Subject Classification:] 37K60
\end{description}
\end{abstract}


\section{Introduction}\label{sec-1}

The $\bar{\partial}$ (Dbar) problem concept was introduced by Beals and Coifman \cite{BC-1980-81,BC-1981-82}
as a generalization of the Riemann-Hilbert problem,
in connection with the inverse scattering transform (IST) scheme for
a kind of generalized Ablowitz-Kaup-Newell-Segur system  (see also \cite{BC-CAPM-1984}).
Then, Dbar  problems were employed by  Ablowitz, Bar Yaacov and Fokas in a series of work
\cite{ABF-SAPM-1983,F-PRL-1983,FA-LNP-1983,FA-JMP-1984}
to solve $(2+1)$-dimensional equations.
In particular, as mentioned in \cite{FA-JMP-1984},
Dbar problem is of crucial importance in solving the Kadomtsev-Petviashvili (KP) II equation using the IST scheme since
the equation (KP II) ``was the first  case the inadequacy of the Riemann-Hilbert
formulation of the IST was discovered.''

The direct linearization (DL) approach was first introduced by
Fokas and Ablowitz in 1981 \cite{FA-PRL-1981}.
As pointed out by the authors (Fokas and Ablowitz) in \cite{FA-JMP-1984},
the DL approach is closely related to the perturbation approach of Rosales \cite{R-SAPM-1978}.
In DL scheme, a linear integral equation plays a central role,
whose solution is used to define eigenfunction as well as the potential function.
It can be verified that the eigenfunction as well as potential function satisfy certain Lax pair
and in this way solutions of the nonlinear equation defined by the Lax pair can be formulated
by the integral equation.
As pointed out in  \cite{FA-PRL-1981}, the verification,
``in sprit, similar to that of Zakharov and Shabat'' in 1974  \cite{ZS-FAA-1974}.
In addition to Lax pairs and solutions, the Gel'fand-Levitan-Marchenko (GLM) equation
can also be constructed from the linear integral equation \cite{FA-PRL-1981}.
The DL scheme and Dbar  problem can be connected via the generalized Cauchy integral formula
(also known as the Cauchy-Pompeiu formula),
with examples  given in \cite{FA-JMP-1984}.
Fokas-Ablowitz's DL scheme was  extended to the  KP equation,
the Benjamin-Ono equation, etc., to find their solutions, e.g.
\cite{AF-1982,AFA-PLA-1983-BO,FA-1982,FA-PLA-1983-KP,SAF-JMP-1984,NVQC-PLA-1982,QNC-PLA-1982}.
Recently, the scheme was also extended to the case of elliptic solitons
and the GLM equations for elliptic solitons were constructed \cite{LSZ-2025}.

Soon after \cite{FA-PRL-1981}, the DL scheme of Fokas and Ablowitz  was developed by
the Dutch group in \cite{NQC-PLA-1983,NQVC-PA-1983,QNCV-PA-1984} with a more flexible machinery.
In the Dutch version nonlinear equations can be constructed together with their solutions,
without a need of checking Lax pairs.
It allows the treatment of discrete as well as continuous equations on one and the same
footing, within one formulism.
Its effectiveness has been proved in particular in the investigation of integrable lattice equations,
e.g. \cite{N-LMP-1985,NCW-LNP-1985,NPCQ-IP-1992,F-JPA-2018,FN-PRSA-2017,FN-JMP-2018,FN-PRSA-2021}.
Recently, an elliptic version of this DL scheme was extended for
discrete integrable systems admitting elliptic solitons \cite{NSZ-CMA-2023}.

The paper \cite{ZS-FAA-1974} in 1974 by Zakharov and Shabat
is considered as an initial work of the so-called dressing method.
The paper was reformulated in \cite{ZM-JETP-1978,ZS-FAA-1979} using the language of the Riemann-Hilbert problem.
In \cite{ZS-FAA-1979} they called the method of \cite{ZS-FAA-1974}  ``vesture method''
and in \cite{ZM-FAA-1985} the ``dressing method''.
Dressing method provides an algorithm for constructing exact solutions
of an evolution equation from its related spectral problem with an initial (seed) solution,
while ``time'' is introduced by a dressing step.
It is ``purely local in the coordinates'' and
``avoids the requirement of decay of solutions of infinity'', as described in \cite{ZS-FAA-1979}.

The combination of dressing method and Dbar problem
was introduced in \cite{ZM-FAA-1985} and \cite{BC-PD-1986}.
Both papers included high dimensional integrable equations as examples.
\cite{ZM-FAA-1985} introduced a connection with the GLM equations and showed how exact solutions were obtained,
while \cite{BC-PD-1986} (see also the review article \cite{BC-IP-1989})
provided clearer steps that how the method could be handled.

Apart from the formulation in \cite{ZM-FAA-1985} and \cite{BC-PD-1986} for the Dbar dressing method,
there is another direct approach based the Dbar problem and the related linear integral equation,
developed by Jaulent and Manna in a sequence of work \cite{JM-IP-1986,JM-EPL-1986,JM-IP-1987,JM-JMP-1987},
which was called the ``spatial transform'' method by the authors (Jaulent and Manna).
They used a Dbar problem together with the associated linear integrable equation
(that is due to the generalized Cauchy integral formula) to define an eigenfunction.
By assuming the eigenfunction satisfies a spectral problem under consideration,
and utilizing asymptotic property of the the eigenfunction, they determined
a formula for the potential function.
For the KdV equation, they introduced a so-called
spectral Wronskian (see \cite{JM-JMP-1987})
which turns out to be an entire function with respect to the spectral parameter.
By this spectral Wronskian they could define a squared eigenfunction symmetry
as well as introduce and deal with time evolution.
Since the recursion operator of the KdV hierarchy is a strong symmetry
which has the squared eigenfunction symmetry as an eigenfunction,
they could determine the whole KdV hierarchy and their solutions.
This method was later described with more details
and illustrated with the KdV hierarchy and Toda hierarchy in \cite{JMM-IP-1988}.
In this method the potential functions are formulated in a more natural way,
and   time evolution is introduced not by ``dressing''
but by using the ``spectral Wronskian'' (cf.\cite{JMM-IP-1988} and \cite{BC-IP-1989});

There are two versions of the DL schemes: the Fokas-Ablowitz version \cite{FA-PRL-1981}
and the Dutch version \cite{NQC-PLA-1983,NQVC-PA-1983,QNCV-PA-1984}.
The later introduced an  infinite matrix structure
whose entries, denoted as $\{S^{(i,j)}\}$, are used to formulate
nonlinear equations that appear as closed forms of certain $S^{(i,j)}$.
For example, the lattice potential KdV (lpKdV) equation, lattice potential modified KdV (lpmKdV) equation
and lattice Schwarzian KdV (lSKdV) equation are formulated by
 $S^{(0,0)}$, $S^{(0,-1)}$ and $S^{(-1,-1)}$, respectively.
Such formulations can also be seen from the by-product of the DL approach,
namely the Cauchy matrix approach \cite{NAH-JPA-2009,ZZ-SAPM-2013}.
We provide a brief introduction about the Cauchy matrix approach in Appendix \ref{app-A} in this paper.
One can refer to \cite{NAH-JPA-2009,ZZ-SAPM-2013} and Chapter 9 of \cite{HJN-book-2016}
for more details of the Cauchy matrix approach.
However, for many important lattice equation, such as
Q3($\delta$) and Q2 equation in the Adler-Bobenko-Suris (ABS) list \cite{ABS-CMP-2003},
one needs to introduce deformed $S^{(i,j)}$ (e.g. adding extra parameter(s) or considering their derivatives
with respect to the parameter(s)) to formulate these equations.
In a recent work \cite{WZZZ-JPA-2024}, eigenfunctions of the Lax pair of the lpKdV equation were studied.
Some equations composed by these  eigenfunctions and their connections with some ABS equations
(e.g. Q3($\delta$), Q2, Q1($\delta$) and H3($\delta$)  equation were revealed.
This progress implies there is deeper insight to be understood behind these eigenfunctions
and  the DL variables (such as $\{S^{(i,j)}\}$, $V(a)$ and $S(a,b)$) in the DL approach or the Cauchy matrix approach.

The purpose of this paper is going to reveal more insight of these connections.
Recalling eigenfunctions also paly  central roles in the  Dbar method,
and the  Dbar method is closely related to the DL approach,
we will develop the Dbar (spatial transform) method to the lattice KdV type equations.
We hope to reveal some connections between the eigenfunctions and the DL variables,
explore a mechanism to introduce extra parameters to  $S^{(i,j)}$
and reveal the relation between the parameter ``$\delta$'' in some ABS equations
and the spectral Wronskian in the Dbar method.

The paper is organized as follows.
First, in Sec.\ref{sec-2}, we briefly review the Dbar problem
and explain how it is related to the Fokas-Ablowitz DL scheme through the Cauchy-Pompeiu integral formula.
Then, in Sec.\ref{sec-3} we develop the Dbar (spatial transform) method to the lpKdV equation,
illustrate how this equation arise from the discrete Dbar approach.
The relation between the eigenfunction of the lpKdV Lax pair and some  $\{S^{(0,j)}\}$ is explored.
Next, in Sec.4, we introduce a parameter $c$ and explain how the lpmKdV and lSKdV equation
can be constructed. After that, in Sec.\ref{sec-5} we construct three ABS equations with $\delta$ terms, namely,
H3$(\delta)$, Q1$(\delta)$ and Q3$(\delta)$, together with their solutions.
We will see how the spectral Wronskians play roles in generating these $\delta$ terms.
Sec.\ref{sec-6} provides explicit formulae for $N$-soliton solutions of
all the lattice equations derived in this paper. In this section
we can elaborate the relations between the lpKdV eigenfunctions and the DL variables.
Concluding remarks are presented in Sec.\ref{sec-7}.
There is an appendix where we sketch the Cauchy matrix approach for some lattice KdV type equations.

\section{The Dbar problem and Cauchy-Pompeiu integral formula}\label{sec-2}

For a function $f(p)$, $p\in \mathbb{C}$, let us introduce the notation
\begin{equation}\label{d-bar}
\bar{\partial}=\partial_{\bar p}=\frac{\partial}{\partial \bar p}
=\frac{1}{2}(\partial_{p_1}+\mathrm{i }\partial_{p_2}),
\end{equation}
where $\mathrm{i}$ is the imaginary unit, $p=p_1+\mathrm{i }p_2\in \mathbb{C}$,
$(p_1,p_2)\in \mathbb{R}^2$ and $\bar p$ stands for the complex conjugate of $p$.
A function $f(p)$ is analytic on a domain $\Omega$ if and only if $\bar{\partial} f(p)=0,~p\in \Omega$.
A $\bar{\partial}$ (Dbar) problem means any problem that consists of finding a function
$\psi(p)$ in the complex plane from a $\bar{\partial}$ equation
involving $\psi(p)$ and  $\bar{\partial}\psi(p)$, e.g.
\begin{equation}\label{d-bar-eq}
\bar{\partial}\psi(p)=\psi(p) R(p),
\end{equation}
where $R(p)$ is some given function.
It is more general than a Riemann-Hilbert problem
where $\psi(p)$ is analytic everywhere in $\mathbb{C}$
except on a contour where it has a jump.
An extension of \eqref{d-bar-eq} is
\begin{equation}\label{d-bar-eqn}
\bar{\partial}\psi(p)=\psi(\omega(p)) R(p),
\end{equation}
which is called a quasi-local $\bar{\partial}$ equation, where
$\omega(p)$ is a function of $p$, e.g. $\omega(p)=-p$.

The generalized Cauchy integral formula
(also called the Cauchy-Pompeiu integral formula, due to Pompeiu \cite{P-1909})
reads (see \cite{AF-book-2021})
\begin{eqnarray}
    \psi(p) = \frac{1}{2\pi {\rm i}} \int_{\partial \Omega} \frac{\psi(\mu)}{\mu - p} \, d\mu + \frac{1}{2\pi {\rm i}} \int_{\Omega} \frac{\bar{\partial}\psi(\mu)}{\mu - p} \, d\mu \wedge d\bar{\mu},
\end{eqnarray}
where  $\partial \Omega$ is  a simple closed contour enclosing the finite region $\Omega$.
Note that the generalized Cauchy integral formula also holds on an affine algebraic Riemann surface \cite{H-2012}.
Thus, finding a solution of the $\bar{\partial}$ equation \eqref{d-bar-eqn}
can be converted to solving the linear integral equation
\begin{eqnarray}\label{CP}
    \psi(p) = \frac{1}{2\pi {\rm i}} \int_{\partial \Omega} \frac{\psi(\mu)}{\mu - p} \, d\mu
    + \frac{1}{2\pi {\rm i}} \int_{\Omega} \frac{\psi(\omega(\mu)) R(\mu)}{\mu - p} \, d\mu \wedge d\bar{\mu}.
\end{eqnarray}

Solutions of the same Dbar equation are usually not unique,
but one can characterize the uniqueness as discussed in \cite{JM-IP-1986}.
We assume $\psi(p)$ has an expansion at infinity:
\begin{equation}
\psi(p)=\sum^N_{j=-\infty} \psi^{(j)} p^j,
\end{equation}
where $\sum^N_{j=0} \psi^{(j)} p^j$ denotes the principle part of $\psi(p)$ at infinity.
Then, on one side, letting  $\Omega\to \mathbb{C}$ in \eqref{CP}, one obtains
\begin{equation}\label{int-2}
    \psi(p) = \sum^N_{j=0} \psi^{(j)} p^j+ \frac{1}{2\pi {\rm i}} \int_{\mathbb{C}}
    \frac{\psi(\omega(\mu)) R(\mu)}{\mu - p} \, d\mu \wedge d\bar{\mu}.
\end{equation}
On the other side, it is natural to assume the homogenous form of the above equation, i.e.
\begin{equation}\label{int-3}
    \psi(p) =  \frac{1}{2\pi {\rm i}} \int_{\mathbb{C}}
    \frac{\psi(\omega(\mu)) R(\mu)}{\mu - p} \, d\mu \wedge d\bar{\mu},
\end{equation}
has always zero solution $\psi(p)=0$.
Thus, for two solution $\psi_1(p)$ and $\psi_2(p)$ of the same Dbar equation \eqref{d-bar-eqn},
if they have same  principle parts  at infinity, then we assume $\psi_1(p)=\psi_2(p)$.
This agrees with the uniqueness of solutions of the integral equation \eqref{int-2}.

In the case of integrable systems, $ \psi(p) $ is usually normalized by setting
$\psi(p) \to 1 $ as $p\to \infty$, i.e.
\[\psi(p) \sim 1+ O({1}/{p}), ~~ ~ p\to \infty.\]
It then follows from \eqref{int-2} that
\begin{equation}\label{int-1}
    \psi(p) = 1+ \frac{1}{2\pi {\rm i}} \int_{\mathbb{C}}
    \frac{\psi(\omega(\mu)) R(\mu)}{\mu - p} \, d\mu \wedge d\bar{\mu}.
\end{equation}
If replacing $\psi(p)$ by $\rho \psi(p)$ where $\rho$ is some plane wave factor,
the above equation takes a form of the linear integral equation employed
by Fokas and Ablowitz in their DL approach \cite{FA-PRL-1981}.
Thus, it is understood that the Dbar problem and the related methods
are closely related to the Fokas-Ablowitz  DL approach.
For some (2+1)-dimensional equations, their DL schemes were also given
when they are solved via the IST scheme based on  Dbar problems \cite{FA-JMP-1984}.

\section{The Dbar approach to the lpKdV equation}\label{sec-3}

In this section, we will extend the Dbar approach to the discrete setting by introducing a quasi-local Dbar problem
with a discrete plane wave factor.
We will   demonstrate that the solution of this Dbar problem satisfies
the Lax pair of the lpKdV equation.
In addition, we will explore the connection between the solution of this Dbar problem
and some $\{S^{(0,j)}\}$.

\subsection{Quasi-local Dbar problem for the lpKdV}\label{sec-3-1}

We adopt a specific quasi-local Dbar problem
\begin{eqnarray}\label{Dp}
    \bar{\partial} \phi\left( p\right) =\phi \left( -p\right) R\left( p\right) , \qquad p\in\mathbb{C},
\end{eqnarray}
with the canonical normalization condition
\begin{equation}\label{cnc}
    \phi  ( p ) =1+p^{-1}\phi^{(-1)}+p^{-2}\phi^{(-2)}+\cdots, \qquad \text{as}  \quad  p\rightarrow \infty.
\end{equation}
Inspired by the DL method \cite{NC-AAM-1995,NQC-PLA-1983,QNCV-PA-1984}
and also the Cauchy matrix approach (see \cite{NAH-JPA-2009} or Appendix \ref{app-A} in this paper),
we adopt the discrete plane wave factor
\begin{equation}\label{dpwf}
    R ( p ) =\left(\frac{\alpha-p}{\alpha+p}\right)^n\left(\frac{\beta-p}{\beta+p}\right)^mR_0 (p),
\end{equation}
where $n, m \in \mathbb{Z}$ are discrete independent variables,
 $\alpha, \beta \in \mathbb{C}$ serve as spacing parameters in $n$ and $m$ direction, respectively,
and $R_0\left(p\right)$ is independent of both $n$ and $m$.
With these settings, all $\{\phi^{(j)}\}$ are functions of $(n,m,\alpha,\beta)$ but independent of $p$.
In addition, for a function $u(n,m)$, we employ the following short hands:
\begin{equation}
u:=u(n,m),\qquad  \widetilde{u}:=u(n+1,m),~~\widehat{u}:=u(n,m+1), \qquad  \widehat{\widetilde{u}}:=u(n+1,m+1).
\end{equation}

To connect the above Dbar problem with the lpKdV equation, we consider the following
linear problems:
\begin{subequations}\label{Lax-LM}
\begin{align}
 &   L\left(p \right)\phi(p)= \left( \alpha+p\right) \widetilde{\widetilde{\phi}} (p)+h\widetilde{\phi}(p)
 +\left( \alpha-p\right)\phi(p),\label{L}\\
 &     M(p)\phi(p)=(\beta+p)\widehat{\phi}(p)-(\alpha+p)\widetilde{\phi}(p)+g\phi(p),\label{M}
\end{align}
\end{subequations}
where $h$ and $g$ are defined by $\phi^{(-1)}$ as the following:
\begin{subequations}\label{hg}
\begin{align}
&  h=\phi^{(-1)}-\widetilde{\widetilde{\phi}}{}^{(-1)}-2\alpha, \label{h}\\
&  g=\alpha-\beta+\widetilde{\phi}^{(-1)}-\widehat{\phi}^{(-1)}. \label{g}
\end{align}
\end{subequations}

In the following we are going to prove
\begin{equation}\label{LM}
L\left(p \right)\phi(p)= 0,~~ M(p)\phi(p)=0
\end{equation}
when $\phi(p)$ is a solution of the Dbar equation \eqref{Dp} with the settings \eqref{cnc} and \eqref{dpwf}.
Since $h$ is independent of \(p\),  we can deduce that
\begin{align*}
\bar{\partial}[L\left(p \right)\phi(p)]&= ( \alpha+p ) \bar{\partial}\widetilde{\widetilde{\phi}} (p)
+h\bar{\partial}\widetilde{\phi}(p)+ ( \alpha-p )\bar{\partial}\phi(p)   \\
&= ( \alpha+p ) \widetilde{\widetilde{\phi}} (-p)\widetilde{\widetilde{R}}(p)
+h\widetilde{\phi}(-p)\widetilde{R}(p) + ( \alpha-p )\phi(-p)R  ( p ).
\end{align*}
By noticing the definition \eqref{dpwf} for the plane wave factor $R(p)$, it is written as
\begin{align}
\bar{\partial}[L\left(p \right)\phi(p)]&=( \alpha-p ) \widetilde{\widetilde{\phi}} (-p)\widetilde{R} ( p)
+h\widetilde{\phi}(-p)\widetilde{R} ( p ) + ( \alpha+p )\phi(-p)\widetilde{R} ( p )      \nonumber\\
&=[L\left(-p \right)\phi(-p) ]\widetilde{ R}\left( p\right),
\end{align}
which means $L\left(p \right)\phi(p)$ is also a solution of the Dbar equation \eqref{Dp}.
Meanwhile, substituting \eqref{cnc} into \eqref{L} yields
\begin{align}
 L\left(p \right)\phi(p)&=( \alpha+p) \widetilde{\widetilde{\phi}} (p)+h\widetilde{\phi}(p)+( \alpha-p )\phi(p)
 \nonumber\\
 &= ( \alpha+p ) \left(1+p^{-1} \widetilde{\widetilde{\phi}}{}^{(-1)}+\dots  \right)
 +h\left(1+p^{-1} \widetilde{\widetilde{\phi}}{}^{(-1)}+\cdots \right)
 +\left( \alpha-p\right)\left(1+p^{-1} \widetilde{\widetilde{\phi}}{}^{(-1)}+\cdots\right) \nonumber\\
 &=\left(h-\phi^{(-1)}+\widetilde{\widetilde{\phi}}{}^{(-1)}+2\alpha\right )+O(p^{-1}).
\end{align}
Thus, when $h$ is defined as in \eqref{h}, we have
\begin{eqnarray}
    L\left(p \right)\phi(p)= O(p^{ -1 })\rightarrow 0, ~~ \ \ p\rightarrow \infty.
\end{eqnarray}
In light of the characterization in Sec.\ref{sec-2} for the uniqueness of the solutions of the Dbar problem,
we conclude that  $ L (p)\phi(p)=0$.
Along the same line we can get $M(p)\phi(p)=0$.

The compatibility of \eqref{LM} can be checked as the following.
Introduce
\begin{eqnarray}
    \psi(p)=(\alpha+p)^n(\beta+p)^m\phi(p),
\end{eqnarray}
by which we rewrite \eqref{LM} into
\begin{subequations}
\begin{align}
    p^2\psi &=  \widetilde{\widetilde{\psi}}+h\widetilde{\psi}+\alpha^2\psi,\\
    \widehat{\psi}&=  \widetilde{\psi}-g\psi,
\end{align}
\end{subequations}
where $h$ and $g$ are given in \eqref{hg}.
The compatibility condition $\widetilde{\widetilde{\widehat{\psi}}}=\widehat{\widetilde{\widetilde{\psi}}}$
gives rise to
\begin{eqnarray}\label{cc}
    (h+\widetilde{g})g=\beta^2-\alpha^2,
\end{eqnarray}
where $\beta^2-\alpha^2$ is the integration constant \cite{WZZZ-JPA-2024,ZPZ-TMP-2020}.
In light of \eqref{hg} and taking
\begin{eqnarray}
    w=\phi^{(-1)},
\end{eqnarray}
\eqref{cc} yields the well known lpKdV equation \cite{NC-AAM-1995,NQC-PLA-1983}
\begin{eqnarray}\label{lpKdV}
    (w-\widehat{\widetilde{w}}-\beta-\alpha)(\widetilde{w}-\widehat{w}+\alpha-\beta)=\beta^2-\alpha^2,
\end{eqnarray}
or H1 equation in the ABS list \cite{ABS-CMP-2003} in terms of
$\mathcal{W}=-w-\alpha n-\beta m$,
\begin{eqnarray}\label{H1}
 (\widehat{\widetilde{\mathcal{W}}}-\mathcal{W})(\widehat{\mathcal{W}}-\widetilde{\mathcal{W}})
 =\beta^2-\alpha^2.
\end{eqnarray}

\subsection{Relation between $\phi^{(i)}$ and $S^{(0,j)}$}\label{sec-3-2}

To reveal the relation between  $\phi^{(i)}$ and $S^{(0,j)}$, we substitute
\eqref{cnc} into the Lax pair \eqref{LM}, which yields
\begin{subequations}
\begin{align}
&    2\alpha+\widetilde{\widetilde{\phi}}{}^{(-1)}+h -\phi^{(-1)}=0,\label{r1}\\
&   \alpha\widetilde{\widetilde{\phi}}{}^{(i)}+\widetilde{\widetilde{\phi}}{}^{(i-1)}
+h\widetilde{\phi}^{(i)}+\alpha\phi^{(i)}-\phi^{(i-1)}=0, \  \ \mathrm{for} \ \ i\leq -1, \label{r2}
\end{align}
\end{subequations}
and
\begin{subequations}
\begin{align}
 &   \beta+\widehat{\phi}^{(-1)}-\alpha-\widetilde{\phi}^{(-1)}+g=0,\label{r3}\\
 &   \beta\widehat{\phi}^{(i)}+\widehat{\phi}^{(i-1)}-\alpha\widetilde{\phi}^{(i)}-\widetilde{\phi}^{(i-1)}
 +g {\phi}^{(i)}=0, \  \ \text{for}  \ \ i\leq -1. \label{r4}
\end{align}
\end{subequations}
\eqref{r1} and \eqref{r3} are nothing but \eqref{hg}.
Inserting \eqref{hg} into \eqref{r2} and \eqref{r4}, we obtain two sets of recurrence relations
for $\{{\phi}^{(i)}\}$:
\begin{subequations}\label{3.19}
\begin{align}
&   \alpha\widetilde{\widetilde{\phi}}{}^{(i)}+\widetilde{\widetilde{\phi}}{}^{(i-1)}
-\left(2\alpha+\widetilde{\widetilde{\phi}}{}^{(-1)} -\phi^{(-1)}\right)
\widetilde{\phi}^{(i)}+\alpha\phi^{(i)}-\phi^{(i-1)}=0,  \label{rr2}\\
&   \beta\widehat{\phi}^{(i)}+\widehat{\phi}^{(i-1)}-\alpha\widetilde{\phi}^{(i)}-\widetilde{\phi}^{(i-1)}
 -\left(\beta+\widehat{\phi}^{(-1)}-\alpha-\widetilde{\phi}^{(-1)}\right) {\phi}^{(i)}=0,  \label{rr4}
\end{align}
\end{subequations}

To reveal the connection with $\{S^{(i,j)}\}$,
we present the following relations:
\begin{subequations}\label{3.20-21}
\begin{align}
& p\widetilde{\widetilde{S}}{}^{(i,j)}-\widetilde{\widetilde{S}}{}^{(i,j+1)}+p S^{(i,j)}+S^{(i,j+1)}
-\left(-\widetilde{\widetilde{S}}{}^{(i,0)}+{S}^{(i,0)}+2p\right)
\widetilde{S}^{(0,j)}=0,\label{3.20}
  \\
&  q\widehat{S}^{(i,j)}-\widehat{S}^{(i,j+1)}-p\widetilde{S}^{(i,j)}+\widetilde{S}^{(i,j+1)}
-\left(-\widehat{S}^{(i,0)}+\widetilde{S}^{(i,0)}+q-p\right) {S}^{(0,j)}=0,
\label{3.21}
\end{align}
\end{subequations}
which are obtained from \eqref{Sij-1-dyna} by
$\widetilde{\eqref{eq:Sij-dyna-1}}+\eqref{eq:Sij-dyna-2}$
and $\eqref{eq:Sij-dyna-3}-\eqref{eq:Sij-dyna-2}$, respectively.
Comparing \eqref{3.19} and \eqref{3.20-21}, we can see the correspondence:
\begin{eqnarray}
    \phi^{(i)} \to S^{(0,-i-1)}, ~~~ \alpha \to -p, ~~~  \beta \to -q.
\end{eqnarray}
Note that this correspondence only covers the case $S^{(0,-i-1)}$ for $i\leq-1$,
the complementary part for $i\geq0$ can be found later.
The complete correspondence will be confirmed   in Sec.\ref{sec-6} (see Remark \ref{rem-4}).

\begin{remark}\label{rem-1}
For the function $\phi(p)$ determined by the Dbar problem \eqref{Dp} together with
the normalization condition \eqref{cnc} and plane wave factor \eqref{dpwf},
in light of the assumptions in Sec.\ref{sec-2} and equation \eqref{int-1},
$\phi(p)$ is uniquely defined by the integral equation
\begin{equation}\label{int-4}
    \phi(p) = 1+ \frac{1}{2\pi {\rm i}} \int_{\mathbb{C}}
    \frac{\phi(-\mu) R(\mu)}{\mu - p} \, d\mu \wedge d\bar{\mu}.
\end{equation}
Thus, corresponding to \eqref{cnc}, we have
\begin{equation}\label{phi-j-int}
      \phi ^{ (j)}=-\frac{1}{2\pi {\rm i}}\int_\mathbb{C} \frac{\phi\left( -\mu \right) R ( \mu  )}
      { \mu^{j+1} }  d\mu \wedge d\bar{\mu },\quad \text{for} \quad j\leq -1.
\end{equation}

\end{remark}

\section{The lpmKdV equation and lSKdV equation}\label{sec-4}

In this section, we proceed to explore how the lpmKdV equation and lSKdV equation
arise from our approach. In particular, we give two formulations for the  lSKdV equation.

\subsection{The lpmKdV equation}\label{sec-4-1}

Assume $c\in \mathbb{C}$ is any finite point where $\phi(p)$ is analytic.
Thus $\phi(p)$ admits an expansion in a neighbourhood of $c$:
\begin{equation}\label{phi-c}
    \phi(p)=\phi_c^{(0)}+\phi_c^{(1)}(p-c)+\phi_c^{(2)}(p-c)^2+\cdots,
\end{equation}
where, in light of the definition \eqref{int-4}, the coefficients are uniquely determined by
\begin{subequations}
\begin{align}
& \phi_c^{(0)}= 1+\frac{1}{2\pi {\rm i}}\int_{\mathbb{C}} \frac{\phi(-\mu ) R( \mu)}{ (\mu-c)}
 d\mu \wedge d\bar{\mu},\\
&    \phi_c^{(l)}=\frac{1}{2\pi {\rm i}}\int_{\mathbb{C}} \frac{\phi ( -\mu ) R ( \mu )}{ (\mu-c)^{l+1} }
d\mu \wedge d\bar{\mu},\  \ \text{for}   \ \ l\geq1.
\end{align}
\end{subequations}
Substituting \eqref{phi-c}  into the Lax pair \eqref{LM}, we obtain the recurrence relations
\begin{subequations}\label{4.3}
\begin{align}
& (\alpha+c)\widetilde{\widetilde{\phi}}_c\!\!{}^{(0)}+h\widetilde{\phi}_c^{(0)}+(\alpha-c)\phi_c^{(0)}=0,
\label{lec1}\\
& (\alpha+c)\widetilde{\widetilde{\phi}}_c{}\!\!^{(i)}+\widetilde{\widetilde{\phi}}_c{}\!\!^{(i-1)}
+h\widetilde{\phi}_c^{(i)}+(\alpha-c)\phi_c^{(i)}-\phi_c^{(i-1)}=0, \quad \text{for} \quad i\geq 1,
\label{lec2}
\end{align}
\end{subequations}
and
\begin{subequations}\label{4.4}
\begin{align}
&    (\beta+c)\widehat{\phi}_c^{(0)}-(\alpha+c)\widetilde{\phi}_c^{(0)}+g\phi_c^{(0)}=0, \label{lec3}\\
& (\beta+c)\widehat{\phi}_c^{(i)}+\widehat{\phi}_c^{(i-1)}-(\alpha+c)\widetilde{\phi}_c^{(i)}
-\widetilde{\phi}_c^{(i-1)}+g\phi_c^{(i)}=0,\quad \text{for} \quad i\geq 1.
\label{lec4}
\end{align}
\end{subequations}
From \eqref{lec1} and \eqref{lec3} we have
\begin{equation}
h=(\alpha+c)\frac{\widetilde{\widetilde{\phi}}_c\!\!{}^{(0)}}{\widetilde{\phi}_c^{(0)}}
+(\alpha-c)\frac{\phi_c^{(0)}}{\widetilde{\phi}_c^{(0)}}, ~~~
g=-(\alpha+c)\frac{\widetilde{\phi}_c^{(0)}}{\phi_c^{(0)}}
+(\beta+c)\frac{\widehat{\phi}_c^{(0)}}{\phi_c^{(0)}},
\end{equation}
which are expressed in terms of  $\phi_c^{(0)}$.
Substituting them into the compatible condition \eqref{cc},
we obtain a quadrilateral lattice equation
\begin{equation}\label{va}
    (\alpha-c)\phi_c^{(0)}\widehat{\phi}_c^{(0)}-(\beta-c)\phi_c^{(0)}\widetilde{\phi}_c^{(0)}
    +(\beta+c)\widehat{\phi}_c^{(0)}\widehat{\widetilde{\phi}}_c\!\!{}^{(0)}
    -(\alpha+c)\widetilde{\phi}_c^{(0)}\widehat{\widetilde{\phi}}_c\!\!{}^{(0)}=0.
\end{equation}
This equation is equivalent to the lattice eigenfunction KdV equation for $\varphi$ in \cite{WZZZ-JPA-2024}
(see Eq.(3.1) in \cite{WZZZ-JPA-2024})
under the transformation
\begin{eqnarray}
    \varphi=(\alpha+c)^n(\beta+c)^m\phi_c^{(0)};
\end{eqnarray}
it also corresponds to the lattice equation for $V(a)$ in \cite{NAH-JPA-2009}
(see the right-hand-side of  Eq.(2.41) in \cite{NAH-JPA-2009}).
While  via another transformation
\begin{subequations}\label{vcAB}
\begin{eqnarray}\label{v-c}
    v_c=\theta(c) \phi_c^{(0)}
\end{eqnarray}
where
\begin{eqnarray}\label{AB}
    \theta(c)=\left(\frac{A}{c-\alpha}\right)^n \left(\frac{B}{c-\beta}\right)^m,
    ~~A^2=c^2-\alpha^2, ~~B^2=c^2-\beta^2,
\end{eqnarray}
\end{subequations}
it gives rise to
\begin{eqnarray}\label{H32}
B\left( v_c\widetilde{v}_c+ \widehat{v}_c\widehat{\widetilde{v}}_c\right)
-A\left(v_c\widehat{v}_c- \widetilde{v}_c\widetilde{\widehat{v}}_c \right)=0,
\end{eqnarray}
which is  the lpmKdV equation \cite{NC-AAM-1995} or the H3$(\delta=0)$ equation in the ABS list \cite{ABS-CMP-2003}.

Finally, equating $h$ in \eqref{r1} and \eqref{lec1} and equating $g$ in \eqref{r3} and \eqref{lec3},
we get a B\"acklund transformation between the lpKdV equation \eqref{lpKdV} and the lpmKdV equation \eqref{va}:
\begin{subequations}
\begin{align}
&   2\alpha+\widetilde{\widetilde{\phi}}{}^{(-1)}-\phi^{(-1)}
=-(\alpha+c)\frac{\widetilde{\widetilde{\phi}}_c\!\!{}^{(0)}}{\widetilde{\phi}_c^{(0)}}
-(\alpha-c)\frac{\phi_c^{(0)}}{\widetilde{\phi}_c^{(0)}},\\
& \beta-\alpha+\widehat{\phi}^{(-1)}-\widetilde{\phi}^{(-1)}
=(\alpha+c)\frac{\widetilde{\phi}_c^{(0)}}{\phi_c^{(0)}}
-(\beta+c)\frac{\widehat{\phi}_c^{(0)}}{\phi_c^{(0)}},
\end{align}
\end{subequations}
which reads in terms of $w$ and $v_c$ as
\begin{eqnarray}
       2\alpha+\widetilde{\widetilde{w}}-w=-\frac{A(\widetilde{\widetilde{v}}_c+v_c)}{\widetilde{v}_c},~~~ \beta-\alpha+\widehat{w}-\widetilde{w}=\frac{-B\widehat{v}_c+A\widetilde{v}_c}{v_c}.
\end{eqnarray}
It connects equations \eqref{lpKdV} and \eqref{H32},
where $A$ and $B$ are defined in \eqref{AB}.

\subsection{The lSKdV equation}\label{sec-4-2}

To formulate the lSKdV equation, we need to consider both $\phi_c^{(0)}$ and $\phi_c^{(1)}$.
In the following we give two ways to derive the lSKdV equation, which will lead  different types of solutions
for this equation.

\subsubsection{The lSKdV with a power background}\label{sec-4-2-1}

Recalling \eqref{4.3} and \eqref{4.4}, from them
we extract those equations  that only involve $\phi_c^{(0)}$ and $\phi_c^{(1)}$, which are
\begin{subequations}\label{4.13}
\begin{align}
&     (\alpha+c)\widetilde{\widetilde{\phi}}_c\!\!{}^{(0)}+h\widetilde{\phi}_c^{(0)}
+(\alpha-c)\phi_c^{(0)}=0,\\
& (\alpha+c)\widetilde{\widetilde{\phi}}_c\!\!{}^{(1)}+\widetilde{\widetilde{\phi}}_c\!\!{}^{(0)}
+h\widetilde{\phi}_c^{(1)}+(\alpha-c)\phi_c^{(1)}-\phi_c^{(0)}=0,
\end{align}
\end{subequations}
and
\begin{subequations}\label{4.14}
\begin{align}
 &   (\beta+c)\widehat{\phi}_c^{(0)}-(\alpha+c)\widetilde{\phi}_c^{(0)}+g\phi_c^{(0)}=0,\\
 &     (\beta+c)\widehat{\phi}_c^{(1)}+\widehat{\phi}_c^{(0)}-(\alpha+c)\widetilde{\phi}_c^{(1)}
 -\widetilde{\phi}_c^{(0)}+g\phi_c^{(1)}=0.
\end{align}
\end{subequations}
Eliminating $h$ from \eqref{4.13} yields
\begin{equation}\label{skdv1}
    (\alpha+c)(\widetilde{\widetilde{\phi}}_c\!\!{}^{(1)}\widetilde{\phi}_c^{(0)}
    -\widetilde{\widetilde{\phi}}_c\!\!{}^{(0)}\widetilde{\phi}_c^{(1)})
    +\widetilde{\widetilde{\phi}}_c\!\!{}^{(0)}\widetilde{\phi}_c^{(0)}- (\alpha-c)(\widetilde{\phi}_c^{(1)}\phi_c^{(0)}-\widetilde{\phi}_c^{(0)}\phi_c^{(1)})
    -\widetilde{\phi}_c^{(0)}\phi_c^{(0)}=0,
\end{equation}
and eliminating $g$ from \eqref{4.14} we obtain
\begin{eqnarray}\label{skdv2}
     (\beta+c)(\widehat{\phi}_c^{(1)}\phi_c^{(0)}-\widehat{\phi}_c^{(0)}\phi_c^{(1)})
     +\widehat{\phi}_c^{(0)}\phi_c^{(0)}- (\alpha+c)(\widetilde{\phi}_c^{(1)}\phi_c^{(0)}-\widetilde{\phi}_c^{(0)}\phi_c^{(1)})
     -\widetilde{\phi}_c^{(0)}\phi_c^{(0)}=0.
\end{eqnarray}
Multiplying both sides of \eqref{skdv1} by  $(\alpha+c)$ and introducing a new variable
\begin{eqnarray}\label{U}
    \Upsilon= (\alpha+c)\left(\widetilde{\phi}_c^{(1)}\phi_c^{(0)}-\widetilde{\phi}_c^{(0)}\phi_c^{(1)}\right)
    +\widetilde{\phi}_c^{(0)}\phi_c^{(0)},
\end{eqnarray}
we can rewrite equation \eqref{skdv1} as
\begin{subequations}\label{BT}
\begin{eqnarray}\label{BT-1}
    (\alpha+c)\widetilde{\Upsilon}-(\alpha-c)\Upsilon=2c\,\widetilde{\phi}_c^{(0)}\phi_c^{(0)}.
\end{eqnarray}
This may serve as a part of B\"acklund transformation between the $\Upsilon$-equation
and the $\phi_c^{(0)}$-equation \eqref{va}.
There should have a $(\widehat{~~}, \beta)$ counterpart, which reads
\begin{eqnarray}\label{BT-2}
    (\beta+c)\widehat{\Upsilon}-(\beta-c)\Upsilon=2c\,\widehat{\phi}_c^{(0)}\phi_c^{(0)},
\end{eqnarray}
\end{subequations}
but it cannot be obtained from \eqref{skdv2}.
Note that from \eqref{skdv2} and \eqref{U} we have an alternative expression for $\Upsilon$:
\begin{eqnarray}\label{U-alt}
    \Upsilon= (\beta+c)\left(\widehat{\phi}_c^{(1)}\phi_c^{(0)}-\widehat{\phi}_c^{(0)}\phi_c^{(1)}\right)
    +\widehat{\phi}_c^{(0)}\phi_c^{(0)},
\end{eqnarray}
which is a symmetric form of \eqref{U}.
Anyway, to derive \eqref{BT-2}, we consider
\begin{eqnarray}
        \left( \beta+p\right) \widehat{\widehat{\phi}} (p)+(h+\widetilde{g}+\widehat{g})\widehat{\phi}(p)
        +\left( \beta-p\right)\phi(p)=0,
\end{eqnarray}
which can be derived from the Lax pair \eqref{LM}.
Inserting the expansion \eqref{phi-c}  we have
\begin{subequations}
\begin{align}
&  (\beta+c)\widehat{\widehat{\phi}}_c\!\!{}^{(0)}+h\widehat{\phi}_c^{(0)}+(\beta-c)\phi_c^{(0)}=0,\\
&    (\beta+c)\widehat{\widehat{\phi}}_c\!\!{}^{(1)}+\widehat{\widehat{\phi}}_c\!\!{}^{(0)}
+h\widehat{\phi}_c^{(1)}+(\beta-c)\phi_c^{(1)}-\phi_c^{(0)}=0.
\end{align}
\end{subequations}
This leads to a relation
\begin{equation}\label{skdv3}
    (\beta+c)(\widehat{\widehat{\phi}}_c\!\!{}^{(1)}\widehat{\phi}_c^{(0)}
    -\widehat{\widehat{\phi}}_c\!\!{}^{(0)}\widehat{\phi}_c^{(1)})
    +\widehat{\widehat{\phi}}_c\!\!{}^{(0)}\widehat{\phi}_c^{(0)}- (\beta-c)(\widehat{\phi}_c^{(1)}\phi_c^{(0)}-\widehat{\phi}_c^{(0)}\phi_c^{(1)})
    -\widehat{\phi}_c^{(0)}\phi_c^{(0)}=0,
\end{equation}
which, combined with \eqref{U-alt}, gives to the second equation \eqref{BT-2}.
Now we have the complete B\"acklund transformation \eqref{BT}.

Note that when $c=0$ one finds from \eqref{BT} that $\Upsilon$ is a constant.
For $c\neq0$, through the identity
\begin{eqnarray}
(\widetilde{\phi}_c^{(0)}\phi_c^{(0)})\widehat{(\widetilde{\phi}_c^{(0)}\phi_c^{(0)})}
=(\widehat{\phi}_c^{(0)}\phi_c^{(0)})\widetilde{(\widehat{\phi}_c^{(0)}\phi_c^{(0)})},
\end{eqnarray}
we have an equation for $\Upsilon$:
\begin{eqnarray}
     (\alpha-c)^2\Upsilon\widehat{\Upsilon}-(\beta-c)^2\Upsilon\widetilde{\Upsilon}
     +(\alpha+c)^2\widetilde{\Upsilon}\widehat{\widetilde{\Upsilon}}
     -(\beta+c)^2\widehat{\Upsilon}\widehat{\widetilde{\Upsilon}}
     +(\beta^2-\alpha^2)(\widetilde{\Upsilon}\widehat{\Upsilon}
     +\Upsilon\widehat{\widetilde{\Upsilon}})=0.
\end{eqnarray}
Under the transformation
\begin{eqnarray}
    \Upsilon=\left(\frac{\alpha-c}{\alpha+c}\right)^n \left(\frac{\beta-c}{\beta+c}\right)^mZ(c),
\end{eqnarray}
it reduces to the lSKdV equation \cite{NC-AAM-1995}
(i.e. Q1$(\delta=0)$ equation in the ABS list \cite{ABS-CMP-2003}):
\begin{eqnarray}\label{lSKdV1}
    (\beta^2-c^2)\big(Z(c)-\widehat{Z}(c)\big)\big(\widetilde{Z}(c)-\widehat{\widetilde{Z}}(c)\big)
    -(\alpha^2-c^2)\big(Z(c)-\widetilde{Z}(c)\big)\big(\widehat{Z}(c)-\widehat{\widetilde{Z}}(c)\big)=0.
\end{eqnarray}

Note once again that in this formulation the parameter $c$ should not be zero.
In this case, one will have solutions with a power background for the lSKdV equation
(cf.\cite{HZ-JPA-2009}).

\subsubsection{The lSKdV with a linear background}\label{sec-4-2-2}

Let us give a second formulation of the lSKdV equation.
First, we rewrite equations \eqref{skdv1}, \eqref{skdv2} and \eqref{skdv3} as the  following forms:
\begin{subequations}
\begin{align}
&    (\alpha+c)\left(\frac{\widetilde{\widetilde{\phi}}_c\!\!{}^{(1)}}
{\widetilde{\widetilde{\phi}}_c\!\!{}^{(0)}} -\frac{\widetilde{\phi}_c^{(1)}}{\widetilde{\phi}_c^{(0)}}\right)
+1-(\alpha-c)\left(\frac{\widetilde{\phi}_c^{(1)}\phi_c^{(0)}}
{\widetilde{\phi}_c^{(0)}\widetilde{\widetilde{\phi}}_c\!\!{}^{(0)}}
-\frac{\phi_c^{(1)}}{\widetilde{\widetilde{\phi}}_c\!\!{}^{(0)}}\right)
=\frac{\phi_c^{(0)}}{\widetilde{\widetilde{\phi}}_c\!\!{}^{(0)}},\\
&     (\beta+c)\left(\frac{\widehat{\phi}_c^{(1)}}{\widehat{\phi}_c^{(0)}}
 -\frac{\phi_c^{(1)}}{\phi_c^{(0)}}\right)
 +1-(\alpha+c) \left(\frac{\widetilde{\phi}_c^{(1)}}{\widehat{\phi}_c^{(0)}}
 -\frac{\widetilde{\phi}_c^{(0)}\phi_c^{(1)}}{\widehat{\phi}_c^{(0)}\phi_c^{(0)}}\right)
 =\frac{\widetilde{\phi}_c^{(0)}}{\widehat{\phi}_c^{(0)}},\\
&     (\beta+c)\left(\frac{\widehat{\widehat{\phi}}_c\!\!{}^{(1)}}{\widehat{\widehat{\phi}}_c\!\!{}^{(0)}}
-\frac{\widehat{\phi}_c^{(1)}}{\widehat{\phi}_c^{(0)}}\right)
+1-(\beta-c)\left(\frac{\widehat{\phi}_c^{(1)}\phi_c^{(0)}}{\widehat{\phi}_c^{(0)}
\widehat{\widehat{\phi}}_c\!\!{}^{(0)}}-\frac{\phi_c^{(1)}}{\widehat{\widehat{\phi}}_c\!\!{}^{(0)}}\right)
=\frac{\phi_c^{(0)}}{\widehat{\widehat{\phi}}_c\!\!{}^{(0)}}.
\end{align}
\end{subequations}
By introducing a new variable
\begin{eqnarray}
    z_c=\frac{\phi_c^{(1)}}{\phi_c^{(0)}},
\end{eqnarray}
the above three equations are rewritten as the following:
\begin{subequations}\label{Tzv}
\begin{align}
&     \frac{1+(\alpha+c)(\widetilde{\widetilde{z}}_c-\widetilde{z}_c)}{1+(\alpha-c)(\widetilde{z}_c-z_c)}
     =\frac{\phi_c^{(0)}}{\widetilde{\widetilde{\phi}}_c\!\!{}^{(0)}}~,\\
&      \frac{1+(\beta+c)(\widehat{z}_c-z_c)}{1+(\alpha+c)(\widetilde{z}_c-z_c)}
=\frac{\widetilde{\phi}_c^{(0)}}{\widehat{\phi}_c^{(0)}}~,\\
&       \frac{1+(\beta+c)(\widehat{\widehat{z}}_c-\widehat{z}_c)}{1+(\beta-c)(\widehat{z}_c-z_c)}
       =\frac{\phi_c^{(0)}}{\widehat{\widehat{\phi}}_c\!\!{}^{(0)}}~.
\end{align}
\end{subequations}
Using the identity
\begin{eqnarray}
    \widehat{\left(\frac{\widetilde{\phi}_c^{(0)}}{\widehat{\phi}_c^{(0)}}\right)}
    \cdot\widetilde{\left(\frac{\widetilde{\phi}_c^{(0)}}{\widehat{\phi}_c^{(0)}}\right)}
    \cdot\frac{\phi_c^{(0)}}{\widetilde{\widetilde{\phi}}_c\!\!{}^{(0)}}
    =\frac{\phi_c^{(0)}}{\widehat{\widehat{\phi}}_c\!\!{}^{(0)}},
\end{eqnarray}
we have a discrete equation
\begin{eqnarray}\label{zc}
      \frac{1+(\beta+c)(\widetilde{\widehat{z}}_c-\widetilde{z}_c)}
      {1+(\alpha+c)(\widetilde{\widehat{z}}_c-\widehat{z}_c)}
      \cdot\frac{1+(\beta-c)(\widehat{z}_c-z_c)}{1+(\alpha-c)(\widetilde{z}_c-z_c)}=1,
\end{eqnarray}
which is the special case of the Nijhoff-Quispel-Capel (NQC) equation (cf.\eqref{NQC-S})
\cite{NQC-PLA-1983,NAH-JPA-2009}.

Note that in this case the derivation allows a degeneration $c=0$.
When $c=0$, under the coordinate transformation
\begin{eqnarray}\label{z}
    z=z_0+\frac{n}{\alpha}+\frac{m}{\beta},
\end{eqnarray}
where $z_0$ denotes $z_c|_{c=0}$,
the transformation \eqref{Tzv} and equation \eqref{zc} reduce to
\begin{subequations}
\begin{align}
&     \frac{\widetilde{\widetilde{z}}-\widetilde{z}}{\widetilde{z}-z}
=\frac{\phi_0^{(0)}}{\widetilde{\widetilde{\phi}}_0\!\!{}^{(0)}},\\
&     \frac{\beta(\widehat{z}-z)}{\alpha(\widetilde{z}-z)}
=\frac{\widetilde{\phi}_0^{(0)}}{\widehat{\phi}_0^{(0)}},\\
&    \frac{\widehat{\widehat{z}}-\widehat{z}}{\widehat{z}-z}
=\frac{\phi_0^{(0)}}{\widehat{\widehat{\phi}}_0\!\!{}^{(0)}},
\end{align}
\end{subequations}
and
\begin{eqnarray}\label{lSKdV2}
    \frac{(\widetilde{\widehat{z}}-\widetilde{z})(\widehat{z}-z)}
    {(\widetilde{\widehat{z}}-\widehat{z})(\widetilde{z}-z)}=\frac{\alpha^2}{\beta^2},
\end{eqnarray}
where $\phi_0^{(0)}=\phi_c^{(0)}|_{c=0}$.
Equation \eqref{lSKdV2} gives the lSKdV equation as well,
whose solution is formulated by \eqref{z} with $z_0$ and a linear background (cf.\cite{HZ-JPA-2009}),
where
\begin{eqnarray}\label{z-0}
    z_0=\frac{\phi_0^{(1)}}{\phi_0^{(0)}}.
\end{eqnarray}

\section{Lattice equations with a $\delta$ term}\label{sec-5}

Some ABS equations contain a $\delta$ term, such as H3$(\delta)$, Q1$(\delta)$
and Q3$(\delta)$, where $\delta$ appears as a parameter.
In \cite{WZZZ-JPA-2024} H3$(\delta)$ and Q3$(\delta)$ were constructed from
the eigenfunctions of the lpKdV Lax pair, while  Q1$(\delta)$ was obtained as a degeneration
of Q3$(\delta)$, as what have been done in \cite{NAH-JPA-2009}.
In the following we will derive these three equations directly from $\phi(p)$ and its deformations.
In addition, we will reveal the connection between the parameter $\delta$ in H3$(\delta)$
and the so-called ``spectral Wronskian'' that played a crucial role
in the Dbar spatial transform method \cite{JM-JMP-1987,JMM-IP-1988}.

\subsection{H3$(\delta)$ equation}\label{sec-5-1}

For the Dbar problem \eqref{Dp}, its spectral Wronskian can be introduced as (cf.\cite{JM-JMP-1987})
\begin{eqnarray}\label{spW}
W(\psi, \phi)= \begin{vmatrix}
\psi(p) &  \phi (p)\\
\psi(-p) &  \phi (-p)
\end{vmatrix},
\end{eqnarray}
where $\psi(p)$ and $\phi (p)$ are two solutions of \eqref{Dp}.
It is easy to check that $\bar{\partial}W(\psi, \phi)=0$, which means
$W(\psi, \phi)$ is an entire function of $p$ on the complex plane.
For $\phi(p)$ being a solution of the Dbar problem \eqref{Dp} with $R(p)$ defined in \eqref{dpwf},
it is easy to see that $(\alpha+p)\widetilde{\phi} (p)$ is also a solution of the same Dbar problem.
These two solutions compose a spectral Wronskian
\begin{eqnarray}\label{spW-Dp}
    D(p)= \begin{vmatrix}
\phi(p) & (\alpha+p)\widetilde{\phi} (p)\\
\phi(-p) & (\alpha-p)\widetilde{\phi} (-p)
\end{vmatrix}.
\end{eqnarray}
Due to the symmetric form of the plane wave factor $R(p)$
(as well as the multi-dimensionally consistent property of the related lattice equations),
the above  $D(p)$ can be equivalently defined as
\begin{eqnarray}\label{spW-Dp-2}
    D(p)=\begin{vmatrix}
\phi(p) & (\beta+p)\widehat{\phi}(p)\\
\phi(-p) & (\beta-p)\widehat{\phi}(-p)
\end{vmatrix}.
\end{eqnarray}
In fact, they are connected by $M(p)\phi(p)=0$ in the Lax pair \eqref{LM}.

Thus we have
\begin{eqnarray}
    \bar{\partial}D(p)=0,
\end{eqnarray}
i.e. $D(p)$ is an entire function of $p$ on the complex plane.
Noticing (from \eqref{cnc})   the asymptotic behavior
\begin{eqnarray}
    D(p)\rightarrow-2p,\quad \text{as} \quad p\rightarrow \infty,
\end{eqnarray}
and also noticing that $D(p)$ is odd with respect to $p$, i.e. $D(p)=-D(-p)$,
we are immediately led to
\begin{eqnarray}\label{delta}
    D(p)=-2p,
\end{eqnarray}
in light of the Liouville Theorem.

Inserting  $p=c$ in $D(p)$, we have
\begin{eqnarray}\label{spW-Dc}
D(c)=    \begin{vmatrix}
\phi(c) & (\alpha+c)\widetilde{\phi} (c)\\
\phi(-c) & (\alpha-c)\widetilde{\phi} (-c)
\end{vmatrix} =-2c,
\end{eqnarray}
and noticing that  $\phi(p=c)=\phi_c^{(0)}$  we obtain
\begin{eqnarray}\label{deltac}
    (\alpha-c)\phi_c^{(0)}\widetilde{\phi}_{-c}^{(0)}- (\alpha+c)\widetilde{\phi}_c^{(0)}\phi_{-c}^{(0)}=-2c.
\end{eqnarray}
Meanwhile, from \eqref{spW-Dp-2} we have
\begin{eqnarray}
     (\beta-c)\phi_c^{(0)}\widehat{\phi}_{-c}^{(0)}- (\beta+c)\widehat{\phi}_c^{(0)}\phi_{-c}^{(0)}=-2c.
\end{eqnarray}

We now construct the H3($\delta$) equation.
First, we define $\mathrm{v}$ through the following linear combination of the H3(0) solution $v_c$:
\begin{align}
    \mathrm{v}(c)&= C_1v_c+C_2v_{-c}\nonumber\\
    &= C_1\theta(c) \phi_c^{(0)}+C_2\theta^{-1}(c)\phi_{-c}^{(0)},
\end{align}
where  $v_c$ is defined by \eqref{vcAB} through $\phi_c^{(0)}$,
and $C_1, C_2$ are constants.
Next, through a calculation  we find the following expression:
\begin{align}
    &\frac{1}{A}\mathrm{v}(c)\widetilde{\mathrm{v}}(c)
    +\frac{1}{A}\widehat{\mathrm{v}}(c)\widetilde{\widehat{\mathrm{v}}}(c)
    -\frac{1}{B}\mathrm{v}(c)\widehat{\mathrm{v}}(c)
    -\frac{1}{B}\widetilde{\mathrm{v}}(c)\widehat{\widetilde{\mathrm{v}}}(c)\nonumber\\
=~ &  \frac{C_1^2\theta^2(c)}{(c-\alpha)(c-\beta)}\Xi_1
    +\frac{C_1C_2}{c^2-\alpha^2}(\Xi_2+\widehat{\Xi}_2)
   -\frac{C_1C_2}{c^2-\beta^2}(\Xi_3+\widetilde{\Xi}_3)
    +\frac{C_2^2\theta^{-2}(c)}{(c+\alpha)(c+\beta)}\Xi_4,
\end{align}
where
\begin{eqnarray*}
    \Xi_1&=&(c-\beta)\phi_c^{(0)}\widetilde{\phi}_c^{(0)}
    +(c+\beta)\widehat{\phi}_c^{(0)}\widehat{\widetilde {\phi}}_c\!\!{}^{(0)}-(c-\alpha)\phi_c^{(0)}\widehat{\phi}_c^{(0)}
    -(c+\alpha)\widetilde{\phi}_c^{(0)}\widehat{\widetilde{\phi}}_c\!\!{}^{(0)},\\
    \Xi_2&=&(c-\alpha)\phi_c^{(0)}\widetilde{\phi}_{-c}^{(0)}
    +(c+\alpha)\widetilde{\phi}_c^{(0)}\phi_{-c}^{(0)},\\
    \Xi_3&=&(c-\beta)\phi_c^{(0)}\widehat{\phi}_{-c}^{(0)}
    +(c+\beta)\widehat{\phi}_c^{(0)}\phi_{-c}^{(0)},\\
    \Xi_4&=&(c+\beta)\phi_{-c}^{(0)}\widetilde{\phi}_{-c}^{(0)}
    +(c-\beta)\widehat{\phi}_c^{(0)}\widehat{\widetilde{\phi}}_{-c}\!\!\!\!\!{}^{(0)}
    -(c+\alpha)\phi_{-c}^{(0)}\widehat{\phi}_{-c}^{(0)}
    -(c-\alpha)\widetilde{\phi}_{-c}^{(0)}\widehat{\widetilde{\phi}}_{-c}\!\!\!\!\!{}^{(0)}.
\end{eqnarray*}
According to \eqref{va} and \eqref{delta}, we have
\begin{eqnarray}
    \Xi_1=0,\quad \Xi_2=-2c,\quad \Xi_3=-2c,\quad \Xi_4=0.
\end{eqnarray}
Thus we arrive at
\begin{eqnarray}\label{H3d}
     \frac{1}{A}\mathrm{v}(c)\widetilde{\mathrm{v}}(c)
     +\frac{1}{A}\widehat{\mathrm{v}}(c)\widehat{\widetilde{\mathrm{v}}}(c)
     -\frac{1}{B}\mathrm{v}(c)\widehat{\mathrm{v}}(c)
     -\frac{1}{B}\widetilde{\mathrm{v}}(c)\widehat{\widetilde{\mathrm{v}}}(c)
     =-4cC_1C_2\left(\frac{1}{A^2}-\frac{1}{B^2}\right),
\end{eqnarray}
where $A, B$ are defined as in \eqref{AB}.
This is the ABS H3$(\delta)$ equation with $\delta=-4cC_1C_2$.

\begin{remark}\label{rem-2}
The derivation highlights the spectral Wronskian \eqref{spW-Dc},
which includes essential interplay between $\phi_c^{(0)}$ and $\phi_{-c}^{(0)}$,
both satisfying the  quadrilateral equation \eqref{va}, in generating the $\delta$ term.
\end{remark}

\subsection{Q1($\delta$) equation}\label{sec-5-2}

In the following we provide a direct formulation for Q1($\delta$), rather than deriving it
as a degeneration from Q3($\delta$) (cf.\cite{NAH-JPA-2009,WZZZ-JPA-2024}).

Recall  the recurrence relation \eqref{lec1}  for $p=-c$ and \eqref{lec2} for $i=1$ and $p=c$, i.e.
\begin{subequations}
\begin{align}
& (\alpha-c)\widetilde{\widetilde{\phi}}_{-c}\!\!\!\!\!{}^{(0)}
+h\widetilde{\phi}_{-c}^{(0)}+(\alpha+c)\phi_{-c}^{(0)}=0,\\
& (\alpha+c)\widetilde{\widetilde{\phi}}_c\!\!{}^{(1)}+\widetilde{\widetilde{\phi}}_c\!\!{}^{(0)}
+h\widetilde{\phi}_c^{(1)}+(\alpha-c)\phi_c^{(1)}-\phi_c^{(0)}=0.
\end{align}
\end{subequations}
Eliminating $h$ from them yields an equation
\begin{align}
   & (\alpha+c)\widetilde{\widetilde{\phi}}_c\!\!{}^{(1)}\widetilde{\phi}_{-c}^{(0)}
   -  (\alpha-c)\widetilde{\phi}_c^{(1)}\widetilde{\widetilde{\phi}}_{-c}\!\!\!\!\!{}^{(0)}
   +\widetilde{\widetilde{\phi}}_{c}\!\!{}^{(0)}\widetilde{\phi}_{-c}^{(0)}
   -  (\alpha+c)\widetilde{\phi}_c^{(1)}\phi_{-c}^{(0)}
   +(\alpha-c)\phi_c^{(1)}\widetilde{\phi}_{-c}^{(0)}-\widetilde{\phi}_{c}^{(0)}\phi_{-c}^{(0)} \nonumber\\
=~& \phi_{c}^{(0)}\widetilde{\phi}_{-c}^{(0)}-\widetilde{\phi}_{c}^{(0)}\phi_{-c}^{(0)}.
\label{q1de1}
\end{align}
Switching $c$ and $-c$ in the above derivation yields another equation
\begin{align}
   & (\alpha-c)\widetilde{\widetilde{\phi}}_{-c}\!\!\!\!\!{}^{(1)}\widetilde{\phi}_{c}^{(0)}
   -  (\alpha+c)\widetilde{\phi}_{-c}^{(1)}\widetilde{\widetilde{\phi}}_{c}\!\!{}^{(0)}
   +\widetilde{\widetilde{\phi}}_{-c}\!\!\!\!\!{}^{(0)}\widetilde{\phi}_{c}^{(0)}
   -  (\alpha-c)\widetilde{\phi}_{-c}^{(1)}\phi_{c}^{(0)}
   +(\alpha+c)\phi_{-c}^{(1)}\widetilde{\phi}_{c}^{(0)}-\widetilde{\phi}_{-c}^{(0)}\phi_{c}^{(0)} \nonumber\\
=~& \phi_{-c}^{(0)}\widetilde{\phi}_{c}^{(0)}-\widetilde{\phi}_{-c}^{(0)}\phi_{c}^{(0)}.
\label{q1de1-2}
\end{align}
Subtracting   \eqref{q1de1-2} from  \eqref{q1de1} gives rise to
\begin{eqnarray}\label{3.2a}
    \widetilde{Y}(c)-Y(c)=2(\phi_{c}^{(0)}\widetilde{\phi}_{-c}^{(0)}-\widetilde{\phi}_{c}^{(0)}\phi_{-c}^{(0)}).
\end{eqnarray}
where we have introduced a new variable
\begin{eqnarray}\label{Vc}
    Y(c)=  (\alpha+c)(\widetilde{\phi}_c^{(1)}\phi_{-c}^{(0)}+\phi_{-c}^{(1)}\widetilde{\phi}_{c}^{(0)})
    -(\alpha-c)(\phi_c^{(1)}\widetilde{\phi}_{-c}^{(0)}+\widetilde{\phi}_{-c}^{(1)}\phi_{c}^{(0)})
    +\widetilde{\phi}_{c}^{(0)}\phi_{-c}^{(0)}-\widetilde{\phi}_{-c}^{(0)}\phi_{c}^{(0)}.
\end{eqnarray}
Note that such a $ Y(c)$  is odd with respect to $c$, i.e.
\begin{eqnarray}\label{Vc-c}
    Y(c)=-Y(-c).
\end{eqnarray}
Next, multiplying \eqref{3.2a} by $(\alpha-c)$ and utilizing the identity \eqref{deltac}, we have
\begin{subequations}
\begin{eqnarray}
        (\alpha-c)\widetilde{Y}(c)-  (\alpha-c)Y(c)+4c=4c\widetilde{\phi}_{c}^{(0)}\phi_{-c}^{(0)}.
\end{eqnarray}
After applying the substitution $c\rightarrow-c$ and employing the antisymmetric property \eqref{Vc-c},
we obtain another relation
\begin{eqnarray}
        (\alpha+c)\widetilde{Y}(c)-  (\alpha+c)Y(c)+4c=4c\widetilde{\phi}_{-c}^{(0)}\phi_{c}^{(0)}.
\end{eqnarray}
\end{subequations}

If we start from the recurrence relation \eqref{lec3} and \eqref{lec4},
in a similar way,  we obtain  another expression of $Y(c)$:
\begin{eqnarray}
    Y(c)=  (\beta+c)(\widehat{\phi}_c^{(1)}\phi_{-c}^{(0)}+\phi_{-c}^{(1)}\widehat{\phi}_{c}^{(0)})
    -(\beta-c)(\phi_c^{(1)}\widehat{\phi}_{-c}^{(0)}+\widehat{\phi}_{-c}^{(1)}\phi_{c}^{(0)})
    +\widehat{\phi}_{c}^{(0)}\phi_{-c}^{(0)}-\widehat{\phi}_{-c}^{(0)}\phi_{c}^{(0)},
\end{eqnarray}
which is equivalent to \eqref{Vc} in light of the Lax pair \eqref{LM}.
In addition, we can also get
\begin{subequations}
\begin{align}
 &       (\beta-c)\widehat{Y}(c)-  (\beta-c)Y(c)+4c=4c\widehat{\phi}_{c}^{(0)}\phi_{-c}^{(0)},\\
 & (\beta+c)\widehat{Y}(c)-  (\beta+c)Y(c)+4c=4c\widehat{\phi}_{-c}^{(0)}\phi_{c}^{(0)}.
\end{align}
\end{subequations}

Now, employing  the identity
\begin{eqnarray}
   \left (\widetilde{\phi}_{c}^{(0)}\phi_{-c}^{(0)}\right)
   \widehat{\left(\widetilde{\phi}_{-c}^{(0)}\phi_{c}^{(0)}\right)}
   =\left(\widehat{\phi}_{c}^{(0)}\phi_{-c}^{(0)}\right)
   \widetilde{\left(\widehat{\phi}_{-c}^{(0)}\phi_{c}^{(0)}\right)},
\end{eqnarray}
we have
\begin{align}
   &\big[ (\alpha-c)\widetilde{Y}(c)-  (\alpha-c)Y(c)
   +4c\big] \big[ (\alpha+c)\widehat{\widetilde{Y}}(c)- (\alpha+c)\widehat{Y}(c)+4c\big] \nonumber\\
   =~&\big[(\beta-c)\widehat{Y}(c)-  (\beta-c)Y(c)
   +4c\big]\big[(\beta+c)\widetilde{\widehat{Y}}(c)-  (\beta+c)\widetilde{Y}(c)+4c\big].
   \label{Q1-1}
\end{align}
This equation can be regarded as Q1(1) in the ABS list.
Through a transformation
\begin{eqnarray}
    2\frac{\mathcal{V}}{\sqrt{\delta}}+\frac{4\alpha c}{c^2-\alpha^2}n+\frac{4\beta c}{c^2-\beta^2}m=Y(c),
\end{eqnarray}
one can introduce $\delta$ and arrive at the Q1$(\delta)$ equation for $\mathcal{V}$
\begin{eqnarray}\label{Q1d}
    (\alpha^2-c^2)(\widetilde{\mathcal{V}}-\mathcal{V})(\widetilde{\widehat{\mathcal{V}}}-\widehat{\mathcal{V}})
    - (\beta^2-c^2)(\widehat{\mathcal{V}}-\mathcal{V})(\widetilde{\widehat{\mathcal{V}}}
    -\widetilde{\mathcal{V}})=4c^2\delta\left(\frac{1}{\alpha^2-c^2}-\frac{1}{\beta^2-c^2}\right).
\end{eqnarray}
Note that in the above formulation for Q1$(\delta)$, it is not allowed to take the parameter $c=0$,
which otherwise leads to $Y(0)=0$ due to \eqref{Vc-c}.

\subsection{NQC and Q3($\delta$) equation}\label{sec-5-3}

In \cite{WZZZ-JPA-2024} the NQC and NQC($\delta$) equation were formulated in terms of
eigenfunctions of the lpKdV Lax pair, then they were extended to the Q3 and Q3($\delta$) equation
in the ABS list.
In the following we will construct these four equation from a  generalized spectral Wronskian.
The construction combines the ideas of   \cite{WZZZ-JPA-2024} and \cite{NAH-JPA-2009}.

\subsubsection{NQC and Q3$(0)$}\label{sec-5-3-1}

Suppose $\phi(p)$ is a solutions of the Dbar problem \eqref{Dp} with the settings \eqref{cnc} and \eqref{dpwf}.
Suppose $\phi(q)$ is a solution of the same problem with the same settings, i.e.
\begin{eqnarray}\label{Dq}
    \bar{\partial} \phi(q ) =\partial_{\bar{q}}\phi(q)=\phi ( -q ) R ( q) , \qquad q\in\mathbb{C}.
\end{eqnarray}
We consider a generalized spectral Wronskian, denoted by $D(p,q)$, defined as (assume  $p^2 \neq q^2$)
\begin{equation}\label{D-pq}
    D(p,q)= \begin{vmatrix}
\phi(p) & (\alpha+p)\widetilde{\phi} (p)\\
\phi(q) & (\alpha+q)\widetilde{\phi} (q)
\end{vmatrix} = (\alpha+q)\phi(p)\widetilde{\phi} (q)-(\alpha+p)\widetilde{\phi} (p)\phi(q).
\end{equation}
Note that apart from the Lax pair \eqref{LM} for $\phi(p)$, we also have
\begin{equation}\label{LM-q}
L\left(q \right)\phi(q)= 0,~~ M(q)\phi(q)=0.
\end{equation}
Making use of the relations $L\left(p \right)\phi(p)= L\left(q \right)\phi(q)= 0$ in the Lax pairs, we find
\begin{eqnarray}
    \widetilde{D}(p,q)= (\alpha-p)\phi(p)\widetilde{\phi} (q)-(\alpha-q)\widetilde{\phi} (p)\phi(q).
\end{eqnarray}
By this relation together with \eqref{D-pq},
one can express $\phi(p)\widetilde{\phi}(q)$ and $\widetilde{\phi} (p)\phi(q)$ as a linear combination of
$D(p,q)$ and $\widetilde{D}(p,q)$:
\begin{eqnarray}
(p^2-q^2)\widetilde{\phi} (p)\phi(q)&=&(\alpha-p)D(p,q)-(\alpha+q)\widetilde{D}(p,q),\\
    (p^2-q^2)\phi(p)\widetilde{\phi}(q)&=&(\alpha-q)D(p,q)-(\alpha+p)\widetilde{D}(p,q).
\end{eqnarray}
Meanwhile, through the Lax pairs \eqref{LM} and \eqref{LM-q},
we have an alternative expression for \eqref{D-pq}, which is
\begin{eqnarray}
    D(p,q)=\begin{vmatrix}
\phi(p) & (\beta+p)\widehat{\phi} (p)\\
\phi(q) & (\beta+q)\widehat{\phi} (q)\\
\end{vmatrix}.
\end{eqnarray}
Similarly, we also have
\begin{eqnarray}
(p^2-q^2)\widehat{\phi} (p)\phi(q)&=&(\beta-p)D(p,q)-(\beta+q)\widehat{D}(p,q),\\
    (p^2-q^2)\phi(p)\widehat{\phi}(q)&=&(\beta-q)D(p,q)-(\beta+p)\widehat{D}(p,q).
\end{eqnarray}
By inserting above equations into the identity
\begin{equation}
    \widetilde{\left(\widehat{\phi} (p)\phi(q)\right)}\phi(p)\widehat{\phi}(q)
    =\widehat{\left(\widetilde{\phi} (p)\phi(q)\right)}\phi(p)\widetilde{\phi}(q),
\end{equation}
we obtain a closed form for $D(p,q)$:
\begin{eqnarray}\label{D}
     (\beta-p)(\beta-q)D(p,q)\widetilde{D}(p,q)-(\alpha-p)(\alpha-q)D(p,q)\widehat{D}(p,q)\nonumber\\
     +(\beta+p)(\beta+q)\widehat{D}(p,q)\widehat{\widetilde{D}}(p,q)
     -(\alpha+p)(\alpha+q)\widetilde{D}(p,q)\widehat{\widetilde{D}}(p,q)\nonumber\\
     +(\alpha^2-\beta^2)\big(\widetilde{D}(p,q)\widehat{D}(p,q)+D(p,q)\widehat{\widetilde{D}}(p,q)\big)=0.
\end{eqnarray}
Via the transformation\footnote{This indicates ${  S}(p,q)={  S}(q,p)$.}
\begin{eqnarray}\label{5.38}
    D(p,q)=(q-p)\big(1-(p+q){  S}(p,q)\big).
\end{eqnarray}
we arrive at the NQC equation (here ${  S}={  S}(p,q)$) \cite{NQC-PLA-1983}
\begin{align}
&\big(1+(\alpha-p){  S}-(\alpha+q)\widetilde{{  S}}\big)
\big(1+(\alpha-q)\widehat{{  S}}-(\alpha+p)\widehat{\widetilde{{  S}}}\big)
\nonumber\\
=& \big(1+(\beta-p){  S}-(\beta+q)\widehat{{  S}}\big)
\big(1+(\beta-q)\widetilde{{  S}}-(\beta+p)\widehat{\widetilde{{  S}}}\big).
\label{NQC-S}
\end{align}
Next, introducing  (cf.\cite{NAH-JPA-2009,HJN-book-2016})
\begin{subequations}\label{X-D}
\begin{equation}
    X(p,q)=\frac{F(p,q)D(p,q)}{q-p},
\end{equation}
where $F(p,q)$ is defined as
\begin{eqnarray}\label{F(p,q)}
    F(p,q)=\left(\frac{\mathcal{A}}{(\alpha-p)(\alpha-q)}\right)^n
    \left(\frac{\mathcal{B}}{(\beta-p)(\beta-q)}\right)^m
\end{eqnarray}
where
\begin{equation}
    \mathcal{A}^2=(\alpha^2-p^2)(\alpha^2-q^2), ~~ \mathcal{B}^2=(\beta^2-p^2)(\beta^2-q^2),
\end{equation}
\end{subequations}
one recovers the Q3$(\delta=0)$ equation \cite{ABS-CMP-2003}
\begin{eqnarray}
    \mathcal{B}\Big(X(p,q)\widetilde{X}(p,q)+\widehat{X}(p,q)\widehat{\widetilde{X}}(p,q)\Big)
    -\mathcal{A}\Big(X(p,q)\widehat{X}(p,q)+\widetilde{X}(p,q)\widehat{\widetilde{X}}(p,q)\Big)\nonumber\\
    +(\alpha^2-\beta^2)\Big(\widetilde{X}(p,q)\widehat{X}(p,q)+X(p,q)\widehat{\widetilde{X}}(p,q)\Big)=0.
    \label{Q3}
\end{eqnarray}

\begin{remark}\label{rem-3}
If we want to express the generalized spectral Wronskian \eqref{D-pq}
in terms of $\phi_c^{(0)}$, we may just take its value at $(p,q)=(c,b)$,
and thus we have
\begin{equation}
    D(c,b) = (\alpha+b)\phi(c)\widetilde{\phi} (b)-(\alpha+c)\widetilde{\phi} (c)\phi(b)= (\alpha+b)\phi_c^{(0)}\widetilde{\phi}_b^{(0)}-(\alpha+c)\widetilde{\phi} _c^{(0)}\phi_b^{(0)}.
\end{equation}
\end{remark}

\subsubsection{Q3$(\delta)$ and NQC$(\delta)$}\label{sec-5-3-2}

We now proceed to construct the Q3$(\delta)$ equation.
Note that in \cite{NAH-JPA-2009} a procedure has been described
that how  Q3$(\delta)$ and NQC equation are related.
The following process is actually similar to the proof presented in \cite{NAH-JPA-2009}.
Here we sketch  the procedure for the completeness of our paper.

Let us first introduce the following function
\begin{equation}\label{u-X}
u=A_1 X(p,q)+A_2 X(p,-q)+A_3 X(-p,q)+A_4 X(-p,-q),
\end{equation}
where $X(p,q)$ is defined as in \eqref{X-D} and $\{A_i\}$ are constants.
For further calculation of the shifts of $u$, we introduce vectors
\begin{eqnarray}\label{r1r2}
    r_1(p)=\left(
\begin{array}{c}
\phi(p) \\
\phi(-p)
\end{array}
\right), ~~ r_2(p)=\left(
\begin{array}{c}
(\alpha+p)\widetilde{\phi}(p) \\
(\alpha-p)\widetilde{\phi}(-p)
\end{array}
\right),
\end{eqnarray}
i.e.
\begin{eqnarray}\label{r1r2-Lambda}
  r_2(p)=\Lambda(p) \widetilde{r}_1(p),
~~~ \Lambda(p) = \left(
\begin{array}{cc}
\alpha+p  & 0 \\
0 &  \alpha-p
\end{array}
\right).
\end{eqnarray}
Using \eqref{r1r2} we can express $u$ as
\begin{eqnarray}\label{u-M}
    u=r_1^T(p) M \,r_2(q)-r_2^T(p) M \,r_1(q),
\end{eqnarray}
where
\begin{eqnarray}
M=\left(
\begin{array}{cc}
    \frac{A_1}{q-p}F(p,q) & \frac{A_2}{-q-p}F(p,-q) \\
    \frac{A_3}{q+p}F(-p,q) & \frac{A_4}{-q+p}F(-p,-q)
\end{array}\right).
\end{eqnarray}
This expression enables us to calculate $\widetilde{u}$.
By using the spectral problem $L\phi=0$ in the Lax pairs \eqref{LM} and \eqref{LM-q}, we can obtain
\begin{eqnarray}\label{u-M-t}
    \widetilde{u}=\frac{\mathcal{A}}{\alpha^2-q^2}r_1^T(p)M\,r_2(q)
    -\frac{\mathcal{A}}{\alpha^2-p^2}r_2^T(p)M\,r_1(q).
\end{eqnarray}
Thus from \eqref{u-M} and \eqref{u-M-t} one can  express $r_1^T(p)M\, r_2(q)$ and $r_2^T(p)M\,r_1(q)$
in terms of  $u$ and $\widetilde{u}$:
\begin{subequations}\label{q3b1b2}
\begin{align}
  &  (q^2-p^2)r_1^T(p)M\, r_2(q)=\mathcal{A}\widetilde{u}-(\alpha^2 -q^2)u, \label{q3b1}\\
  &   (q^2-p^2)r_2^T(p)M\, r_1(q)=\mathcal{A}\widetilde{u}-(\alpha^2 -p^2)u. \label{q3b2}
\end{align}
\end{subequations}

In the next step, we consider the introduction of the $\delta$ term.
To this end, we introduce the following determinant
\begin{align}
    \left|\begin{array}{cc}
       r_1^t(p)Mr_2(q)  & r_1^t(p)Mr_1(q) \\
     r_2^t(p)Mr_2(q)   & r_2^t(p)Mr_1(q)
    \end{array}\right|
= \left|\left(\begin{array}{c}
   r_1^T(p) \\
   r_2^T(p)
\end{array}\right)M \left(r_2(q), r_1(q)\right)\right|
=-4pq\left|M\right|,\label{deltaq3}
\end{align}
where we have made use of the fact (cf.\eqref{spW} and \eqref{delta})
\begin{equation}
\left|r_1(p), r_2(p)\right|=D(p)=-2p.
\end{equation}
Note here that $|M|$ is also a constant
\begin{equation}
|M|=-\frac{A_1 A_4}{(q-p)^2}+\frac{A_2A_3}{(q+p)^2}.
\end{equation}

Among the elements in the above determinant, we define
\begin{eqnarray}\label{U-r1}
    U=r_1^T(p)M\, r_1(q).
\end{eqnarray}
Noticing the relation $r_2(q)=\Lambda(q)\widetilde{r}_1(q)$ and
$\mathcal{A}\widetilde{M}=\Lambda(p)M\Lambda(q)$,
we can find
\begin{eqnarray}\label{U-r2}
\mathcal{A}  \widetilde{U}=r_2^T(p)M\, r_2(q).
\end{eqnarray}
Thus, from the determinant \eqref{deltaq3} we have
\begin{equation}\label{deter1}
   U\widetilde{U}=\frac{1}{(q^2-p^2)^2} \left[\mathcal{A}\widetilde{u}^2-(2\alpha^2-p^2-q^2)u\widetilde{u}+\mathcal{A}u^2\right]
   +\frac{4pq\left|M\right|}{\mathcal{A}}.
\end{equation}
Next, we establish a second relation between $U$ and $u$.
Based on $L\phi=0$ in the Lax pair \eqref{LM-q}, we have
\begin{eqnarray}
    \left(\begin{array}{c}
        (\alpha+q)^2\widetilde{\widetilde{\phi}}(q) \\
(\alpha-q)^2\widetilde{\widetilde{\phi}}(-q)
    \end{array}\right)+h \left(\begin{array}{c}
        (\alpha+q)\widetilde{\phi}(q) \\
(\alpha-q)\widetilde{\phi}(-q)
    \end{array}\right)+ \left(\begin{array}{c}
        (\alpha^2-q^2)\phi(q) \\
(\alpha^2-q^2)\phi(-q)
    \end{array}\right)=0,
\end{eqnarray}
i.e.
\begin{equation}
\Lambda \widetilde{ r}_2(q) + h\, r_2(q) +(\alpha^2-q^2) r_1(q)=0.
\end{equation}
By  multiplying $r_2^T(p) M$ from the left
we obtain
\begin{eqnarray}
    \mathcal{A}\widetilde{\Big(r_1^T(p)M\, r_2(q)\Big)}
    +h\mathcal{A}\widetilde{\Big(r_1^T(p)M\,r_1(q)\Big)}+(\alpha^2-q^2)r_2^T(p)M\,r_1(q)=0.
\end{eqnarray}
Substituting  \eqref{q3b1b2} and \eqref{U-r1}, we arrive at
\begin{eqnarray}\label{U1}
    \mathcal{A}\widetilde{\widetilde{u}}-\mathcal{A}u=h(p^2-q^2)\widetilde{U}.
\end{eqnarray}
Similarly, from $M\phi = 0$  in the Lax pair \eqref{LM-q}, we can derive a relation
\begin{eqnarray}\label{U2}
    \mathcal{A}\widetilde{u}-\mathcal{B}\widehat{u}-(\alpha^2-\beta^2)u=g(q^2-p^2)U.
\end{eqnarray}
Subtracting \eqref{U1} from the tilde-shifted version of \eqref{U2}, we obtain
\begin{eqnarray}\label{U3}
      \mathcal{A}u-\mathcal{B}\widetilde{\widehat{u}}-(\alpha^2-\beta^2)\widetilde{u}
      =(\widetilde{g}+h)(q^2-p^2)\widetilde{U}.
\end{eqnarray}
To get rid of  $h$ and $g$, we take the product of \eqref{U2} and \eqref{U3}, which gives
\begin{eqnarray}\label{q3d_1}
    \big(  \mathcal{A}\widetilde{u}-\mathcal{B}\widehat{u}-(\alpha^2-\beta^2)u\big)
    \big( \mathcal{A}u-\mathcal{B}\widetilde{\widehat{u}}-(\alpha^2-\beta^2)\widetilde{u}\big)
    &=&g(\widetilde{g}+h)(q^2-p^2)^2U\widetilde{U}\nonumber\\
    &=&(\beta^2-\alpha^2)(q^2-p^2)^2U\widetilde{U},
\end{eqnarray}
where we have made use of the compatibility result \eqref{cc}.
Now, combining the above equation and \eqref{deter1} to eliminate $U\widetilde{U}$,
we can obtain a closed form for $u$:
\begin{eqnarray}\label{Q3d}
     \mathcal{B}(u\widetilde{u}+\widehat{u}\widetilde{\widehat{u}})-\mathcal{A}(u\widehat{u}
     +\widetilde{u}\widetilde{\widehat{u}})=(\beta^2-\alpha^2)
     \big(\widetilde{u}\widehat{u}+u\widetilde{\widehat{u}}+\frac{\delta}{\mathcal{A}\mathcal{B}}\big),
 \end{eqnarray}
which is Q3$(\delta)$ equation with
\begin{eqnarray}
    \delta=4pq(q^2-p^2)^2\left|M\right|=-4pq\left((p+q)^2A_1A_4-(p-q)^2A_2A_3\right).
\end{eqnarray}

Finally, from the function $u$ defined in \eqref{u-X}, we introduce a new $\mathcal{S}=\mathcal{S}(p,q)$ by
\begin{eqnarray}
   u=F(p,q)\big(1-(p+q)\mathcal{S}\big),
\end{eqnarray}
which maps the above Q3($\delta$) equation  to the NQC($\delta$) equation \cite{WZZZ-JPA-2024}:
\begin{align}
&    \big(1+(\alpha-p)\mathcal{S}-(\alpha+q)\widetilde{\mathcal{S}}\big)
    \big(1+(\alpha-q)\widehat{\mathcal{S}}-(\alpha+p)\widetilde{\widehat{\mathcal{S}}}\big)\\
&~~~    -\big(1+(\beta-p)\mathcal{S}-(\beta+q)\widehat{\mathcal{S}}\big)
    \big(1+(\beta-q)\widetilde{\mathcal{S}}-(\beta+p)\widetilde{\widehat{\mathcal{S}}}\big)=C,
\label{NQCd}
\end{align}
where
\begin{eqnarray}
    C=\frac{\delta(\alpha^2-\beta^2)F^2(p,q)}{(q+p)^2(\alpha+p)(\alpha+q)(\beta+p)(\beta+q)}.
\end{eqnarray}

\section{$N$-soliton solutions}\label{sec-6}

In this section we present explicit formulae for $ \phi ^{(l)}$ and $\phi_c^{(0)}$, etc.,
from which we can see how they are related to the functions $\{S^{(i,j)}\}$
and $V(p)$ in the DL approach or the Cauchy matrix approach.

Recalling the integral equation \eqref{int-4} related to the Dbar problem \eqref{Dp}, i.e.
\begin{equation}\label{chi}
    \phi \left( p\right) = 1+\frac{1}{2\pi {\rm i}}\int_\mathbb{C}
    \frac{\phi ( -\mu) R(\mu) }{\mu -p}{\rm d}\mu \wedge {\rm d}\bar{\mu },
\end{equation}
where
\begin{equation}\label{R}
    R\left( p\right) =\left(\frac{\alpha-p}{\alpha+p}\right)^n\left(\frac{\beta-p}{\beta+p}\right)^mR_0 (p ).
\end{equation}
This yields the  following expressions (see \eqref{phi-j-int}):
\begin{eqnarray}
      \phi ^{(l)}&=&-\frac{1}{2\pi {\rm i}}
      \int_\mathbb{C} \frac{\phi\left( -\mu \right) R\left( \mu \right)}{ \mu^{l+1} }  d\mu \wedge d\bar{\mu },
      \quad \text{for} \quad l\leq-1, \label{6.3}\\
   \phi_c^{(0)}&=&1+\frac{1}{2\pi {\rm i}}
   \int _{\mathbb{C}}\frac{\phi\left( -\mu \right) R( \mu)}{ (\mu-c)}  d\mu \wedge d\bar{\mu}, \label{6.4}\\
    \phi_c^{(l)}&=&\frac{1}{2\pi {\rm i}}
    \int _{\mathbb{C}}\frac{\phi( -\mu ) R( \mu )}{ (\mu-c)^{l+1} }  d\mu \wedge d\bar{\mu},
    \quad \text{for} \quad l\geq 1. \label{6.5}
\end{eqnarray}

In the Dbar approach, there exists a map from $R_0(p)$ to the exact solutions of the investigated integrable equations.
To construct a $N$-soliton solution, we choose the Dirac-$\delta$ function
\begin{equation}\label{R0-p}
     R_0 ( p ) = 2\pi \sum ^{N}_{j=1}\rho^{(0)}_j \delta \left( p+p_j\right), ~~~~
     p_j, \rho^{(0)}_j \in \mathbb{C}.
\end{equation}
Note that the Dirac-$\delta$ function in the complex plane satisfies the following property
for any sufficiently smooth function $f(z)$:
\begin{eqnarray}
    \int_{\Omega} f(z) \, \delta(z - z_0) \, {\rm d}z\wedge{\rm d}\bar{z} = f(z_0),
\end{eqnarray}
where the domain $\Omega$ contains the point $z_0$.

Substituting \eqref{R} with \eqref{R0-p}  into \eqref{chi} we have
\begin{eqnarray}
\phi \left( p\right)
&=&1+\frac{1}{2\pi {\rm i}}\int_\mathbb{C} \frac{\phi(-\mu )
\left(\frac{\alpha-\mu}{\alpha+\mu}\right)^n\left(\frac{\beta-\mu}{\beta+\mu}\right)^m
2\pi\sum ^{N}_{j=1}\rho^{(0)}_j \delta( \mu +p_{j})}{\mu-p}{\rm d}\mu \wedge {\rm d}\bar{\mu } \nonumber \\
&=&  1-\sum ^{N}_{j=1}\frac{\phi (p_j) \rho_j}{{\rm i}\left(p_j+p\right)},
\label{6.8}
\end{eqnarray}
where
\begin{equation}\label{pwf-rho}
\rho_j=\left(\frac{\alpha+p_j}{\alpha-p_j}\right)^n\left(\frac{\beta+p_j}{\beta-p_j}\right)^m \rho^{(0)}_j.
\end{equation}
Then, taking $p=p_1, p_2, \cdots, p_N$ in \eqref{6.8}, we obtain an equation set
for $\{\phi ( p_{i})\}$:
\begin{eqnarray}
    \phi ( p_{i})+\sum ^{N}_{j=1}\frac{\phi (p_j)\rho_j}{{\rm i}\left(p_j+p_i\right) }=1,
    ~~~ (i=1,2,\cdots,N),
\end{eqnarray}
which can be rewritten as the matrix form
\begin{eqnarray}
    (\bI+\bM)\Phi=\bs,
\end{eqnarray}
where $\bI$ is the $N$th order identity matrix,
\begin{eqnarray}
    \bM=\left(M_{ij}\right)_{N\times N}=\left(\frac{\rho_j}{{\rm i} (p_j+p_i )} \right)_{N\times N}, ~~~
    \Phi=\left(\begin{array}{c}
    \phi(p_1)   \\
     \phi(p_2)  \\
    \vdots \\
     \phi(p_N)
\end{array}\right),~~~
\bs=\left(\begin{array}{c}
     1   \\
    1  \\
    \vdots \\
    1
\end{array}\right).
\end{eqnarray}
Thus we have
\begin{equation}\label{Phi}
\Phi=  (\bI+\bM)^{-1}\bs.
\end{equation}
In addition, $\phi(p)$ given by \eqref{6.8} can be written as
\begin{equation}
    \phi(p)=1+{\rm i}\br^T (\bK-p\bI)^{-1} \left(\bI+\bM\right)^{-1}\bs, \label{phi-pp}
\end{equation}
where
\begin{equation}
\bK=\mathrm{diag}\{-p_1,-p_2,\cdots, -p_N\},~~~
\br=(    \rho_1,     \rho_2,   \cdots,     \rho_N)^T
\end{equation}
with $\rho_j$ being defined in \eqref{pwf-rho}.

For $\phi ^{(l)}$ defined in \eqref{6.3}, by inserting  $ R_0 ( p )$ we obtain
\begin{equation}
\phi ^{(l)}=-{\rm i}\sum ^{N}_{j=1}\phi (p_j ) \rho_j (-p_j)^{-(l+1)},
\end{equation}
which, by means of \eqref{Phi}, can be written into a compact form
\begin{equation}\label{phi-l}
\phi ^{(l)}=-{\rm i}\br^T \bK^{-(l+1)} \left(\bI+\bM\right)^{-1}\bs,~~~ (l\leq -1).
\end{equation}
In addition, repeating the similar procedure for \eqref{6.4} and \eqref{6.5}, we can obtain
\begin{eqnarray}
     \phi_c^{(0)}&=&1+{\rm i}\br^T (\bK-c\bI)^{-1} \left(\bI+\bM\right)^{-1}\bs, \label{phi-c0}\\
    \phi_c^{(l)}&=&{\rm i}\br^T (\bK-c\bI)^{-(l+1)} \left(\bI+\bM\right)^{-1}\bs,\  \ \mathrm{for}  \ \ l\geq1.\label{phi-cl}
\end{eqnarray}

\begin{remark}\label{rem-4}
The formulae \eqref{phi-l}, \eqref{phi-c0} and \eqref{phi-cl} indicate their correspondences to the master functions
in the DL approach or the Cauchy matrix approach, which are (cf.\eqref{Sij-1} and \eqref{Va})
\begin{subequations}
\begin{align}
& \phi ^{(l)} \to S^{(0,-l-1)},~~~ (l\leq -1),\\
& \phi_0^{(0)}-1 \to S^{(0,-1)},\\
& \phi_0^{(l)}\to S^{(0,-l-1)},~~~ (l \geq 1),\\
& \phi_c^{(0)} \to V(c),\\
& \phi(p) \to V(p).
\end{align}
\end{subequations}
\end{remark}

Finally, we summarize the $N$-soliton formulae of the discrete integrable equations
 we have derived in this paper, which are expressed in terms of the coefficients
 $\{\phi^{(l)}\}$ and $\{\phi^{(l)}_{c}\}$, as presented in
 Table \ref{tab:my_table}.

\begin{table}[H]

\begin{tabular}{|l|l|>{\raggedright\arraybackslash}p{0.6\linewidth}|}
\hline
{equation}   & {variable} &   {solution} \\[0.7ex]
\hline
lpKdV \eqref{lpKdV} & $w$ &     $\phi^{(-1)}$  \\[0.7ex]
\hline
H1 \eqref{H1} &$\mathcal{W}$&  $-\phi^{(-1)}-\alpha n-\beta m$  \\[0.7ex]
\hline
Eq.\eqref{va}    &$ \phi_c^{(0)} $&   $ \phi_c^{(0)} $ \\[0.7ex]
\hline
 lpmKdV \eqref{H32} & $v_c$ &   $\theta(c) \phi_c^{(0)}$\\[0.7ex]
\hline
lSKdV \eqref{lSKdV1}  &$Z(c)$&    $(\frac{\alpha-c}{\alpha+c})^{-n}(\frac{\beta-c}{\beta+c})^{-m}
\big( (\alpha+c)(\widetilde{\phi}_c^{(1)}\phi_c^{(0)}-\widetilde{\phi}_c^{(0)}\phi_c^{(1)})
+\widetilde{\phi}_c^{(0)}\phi_c^{(0)} \big)$\\[0.7ex]
\hline
lSKdV  \eqref{lSKdV2}      &$z$&     $\phi_0^{(1)}/\phi_0^{(0)}+n/\alpha+m/\beta$ \\[0.7ex]
\hline
H3($\delta$) \eqref{H3d}      &$\mathrm{v}(c)$&     $C_1\theta(c) \phi_c^{(0)}+C_2\theta^{-1}(c)\phi_{-c}^{(0)}$ \\[0.7ex]
\hline
 Q1($\delta$) \eqref{Q1d}&$\mathcal{V}(c)$ &   $\Theta$ \\
 \hline
Q3($0$) \eqref{Q3}   & $X(p,q)$ &     $\frac{F(p,q)}{q-p}D(p,q)$\\
\hline
NQC \eqref{NQC-S}   & $S(p,q)$&     $\frac{1}{p+q}(1-\frac{D(p,q)}{q-p})$\\
\hline
Q3($\delta$) \eqref{Q3d}   & $u$ &     $\Xi$\\
\hline
NQC($\delta$) \eqref{NQCd}   & $\mathcal{S}(p,q)$&     $\frac{1}{p+q}\big(1-\frac{\Xi}{F(p,q)}\big)$\\
\hline
\end{tabular}
\vspace{12pt}
\centering
\caption{Solutions of discrete integrable equations}

\label{tab:my_table}
\end{table}

Notations involved in this table are given below:
\begin{align*}
\phi(p) & =1+{\rm i}\br^T (\bK-p\bI)^{-1} \left(\bI+\bM\right)^{-1}\bs, \\
\phi^{(-1)}&=-{\rm i}\br^T  \left(\bI+\bM\right)^{-1}\bs, \\
\phi_0^{(0)}&=1+{\rm i}\br^T \bK^{-1} \left(\bI+\bM\right)^{-1}\bs, \\
\phi_c^{(0)}&=1+{\rm i}\br^T (\bK-c\bI)^{-1} \left(\bI+\bM\right)^{-1}\bs, \\
\phi_0^{ (1 )}&={\rm i}\br^T \bK^{-2} \left(\bI+\bM\right)^{-1}\bs,\\
\phi_c^{(1)}&={\rm i}\br^T (\bK-c\bI)^{-2} \left(\bI+\bM\right)^{-1}\bs,\\
\Theta &=\frac{\sqrt{\delta}}{2}[(\alpha+c)(\widetilde{\phi}_c^{(1)}\phi_{-c}^{(0)}
+\phi_{-c}^{(1)}\widetilde{\phi}_{c}^{(0)})-(\alpha-c)(\phi_c^{(1)}\widetilde{\phi}_{-c}^{(0)}
+\widetilde{\phi}_{-c}^{(1)}\phi_{c}^{(0)})+\widetilde{\phi}_{c}^{(0)}\phi_{-c}^{(0)}
-\widetilde{\phi}_{-c}^{(0)}\phi_{c}^{(0)}]\\
&~~- \frac{2\alpha c \sqrt{\delta}}{c^2-\alpha^2}n - \frac{2\beta c \sqrt{\delta}}{c^2 - \beta^2}m,\\
\Xi &=A_1 \frac{F(p,q)}{q-p}D(p,q)+A_2 \frac{F(p,-q)}{-q-p}D(p,-q)+A_3 \frac{F(-p,q)}{q+p}D(-p,q)\\
&  ~~ +A_4 \frac{F(-p,-q)}{-q+p}D(-p,-q),\\
D(p,q)&= (\alpha+q)\phi(p)\widetilde{\phi} (q)-(\alpha+p)\widetilde{\phi} (p)\phi(q).
\end{align*}

\section{Concluding remarks}\label{sec-7}

The purpose of this paper is to bridge the gap between the Dbar method and the
DL approach for lattice KdV type equations.
This has been achieved through formulating these equations in the Dbar method
and investigating the relations between the eigenfunction (and its expansions)
and the DL variables $S^{(i,j)}$ and $V(p)$.

We have extended the Dbar (spatial transform) method (introduced by Jaulent and Manna in 1980s)
to the lpKdV equation \eqref{lpKdV}.
We introduced the Dbar problem \eqref{Dp},
accompanied with the asymptotic assumption \eqref{cnc} and the discrete plane wave factor \eqref{dpwf},
which defines the eigenfunction $\phi(p)$ of the Lax pair \eqref{LM} of the lpKdV equation.
On the other hand, in light of  the Cauchy-Pompeiu integral formula, solving the Dbar problem \eqref{Dp}
can be converted to solving the linear integral equation \eqref{int-4}.
The later is the starting point as well as the central role of the Fokas-Ablowitz DL approach (cf.\cite{FA-PRL-1981}).
As usual, expansion of the eigenfunction $\phi(p)$ at infinity can lead to the construction of the lpKdV equation,
which has been described in Sec.\ref{sec-3}.
However, as we have shown in Sec.\ref{sec-4}, expansion of  $\phi(p)$ at a finite point $c$ can
give rise to the lpmKdV equation and the lSKdV equation.
Moreover, we have presented in Sec.\ref{sec-4-2-2} a new formulation for the lSKdV equation,
which seems not have been reported in the DL approach or the Cauchy matrix approach.
To obtain the ABS equations with $\delta$-terms, namely,
the H3$(\delta)$, Q1$(\delta)$ and Q3$(\delta)$ equations,
we have made use of the so-called spectral Wronskian and its double parameter generalization
(see $D(p,q)$ in \eqref{D-pq}).
They enable us to introduce the $\delta$-terms.
We have also shown how the NQC equation and the NQC$(\delta)$ equation arose from our approach,
which corresponds to the formulation via eigenfunctions given in \cite{WZZZ-JPA-2024}.
More than that, in our construction, Q1$(\delta)$ is obtained through a direct derivation,
rather than via a degeneration from Q3$(\delta)$ (cf.\cite{NAH-JPA-2009,WZZZ-JPA-2024}).
In Sec.\ref{sec-6} we have presented explicit formulae for $N$-soliton solutions
for the derived discrete equations.
These formulae enable us to elaborate the connection between $\phi(p)$ (and its expansions at infinity
and at a finite point $c$) and the DL variables $S^{(i,j)}$ and $V(c)$.
Their correspondences are pointed out in Remark \ref{rem-4} in Sec.\ref{sec-6}.

In this paper, we derived the NQC equation that was formulated by $S(a,b)$ in the
DL approach or the Cauchy matrix approach (see \eqref{NQC}).
The formulation of $S(a,b)$ in our paper is given by \eqref{5.38} together with $D(p,q)$
defined in \eqref{D-pq} by $\phi(p)$ and $\phi(q)$.
Although $\phi(p)$ can be formulated as in \eqref{phi-pp}, corresponding to
the DL variable $V(p)$ (see \eqref{Va}),
we cannot recover  $S(p,q)$ from $D(p,q)$ together with \eqref{phi-pp}.
However, we would like to point out that $S(p,q)$ defined as in \eqref{Sab-1}
can be formulated from a new Dbar problem, which will be explored in a follow-up paper.
In the future, we will also extend our investigation to the Dbar method for the lattice Boussinesq equations
and the lattice KP type equations,
in particular, to pursue its potential utility in seeking the lattice Boussinesq $\delta$-type equations.

\vskip 20pt
\subsection*{Data availability}

No data was used for the research described in the article.

\subsection*{Conflict of  interests}

The authors declare that they have no financial and non-financial conflict of interests about the work reported in this paper.

\section*{Acknowledgements}

This project is supported by the NSFC grant (Nos. 12271334  and 12171306).

\appendix

\section{Cauchy matrix approach to the lattice KdV type equations}\label{app-A}

In what follows we sketch how the lpKdV, lpmKdV, lSKdV and NQC equations
arise as closed forms from the Cauchy matrix approach.
One can refer to \cite{NAH-JPA-2009} or Chapter 9 of \cite{HJN-book-2016} for details.
We adopt the notations used in \cite{NAH-JPA-2009}.

We start from a Sylvester equation
\begin{equation}
\bM \bK+ \bK \bM = {\br}\bc^T,
\label{rel-MK}
\end{equation}
where
\begin{equation}
\bK=\mathrm{diag}(k_1,k_2,\cdots,k_N),~~
\br=(\rho_1,\rho_2,\cdots,\rho_N)^T, ~~
\bc=(c_1,c_2,\cdots,c_N)^T,
\end{equation}
$\rho_i$ is  the plane wave factor defined as
\begin{equation}
\rho_i=\biggl(\frac{p +k_i}{p -k_i}\biggr)^n\biggl(\frac{q + k_i}{q-k_i}\biggr)^m \rho^0_{i},
\end{equation}
$c_i$, $k_i$ and $\rho^0_{i}$ are constants, $\bM$ is a``dressed'' Cauchy matrix
\begin{equation}
\bM=(M_{i,j})_{N\times N},~~M_{i,j}=\frac{\rho_i c_j}{k_i+k_j}.
\end{equation}

Define the master functions
\begin{equation}
S^{(i,j)}= \bc^T \,\bK^j(\bI+ \bM)^{-1} \bK^i \br, ~~ ~ i,j\in \mathbb{Z},
\label{Sij-1}
\end{equation}
which compose an infinite symmetric matrix, i.e.  (also see \cite{ZZ-SAPM-2013})
\begin{equation}
S^{(i,j)}=S^{(j,i)},~~ ~ i,j\in \mathbb{Z}.
\label{Sij-1-sym}
\end{equation}
It can be proved that $\{S^{(i,j)}\}$ obey dynamical recurrence relations:
\begin{subequations}
\begin{align}
& p \wt{S}^{(i,j)}-\wt{S}^{(i,j+1)}=p S^{(i,j)}+S^{(i+1,j)}-\wt{S}^{(i,0)}S^{(0,j)},\label{eq:Sij-dyna-1}\\
& p S^{(i,j)}+S^{(i,j+1)}=p \wt{S}^{(i,j)}-\wt{S}^{(i+1,j)}+S^{(i,0)}\wt{S}^{(0,j)},\label{eq:Sij-dyna-2}\\
& q \wh{S}^{(i,j)}-\wh{S}^{(i,j+1)}=q S^{(i,j)}+S^{(i+1,j)}-\wh{S}^{(i,0)}S^{(0,j)},\label{eq:Sij-dyna-3}\\
& q S^{(i,j)}+S^{(i,j+1)}=q
\wh{S}^{(i,j)}-\wh{S}^{(i+1,j)}+S^{(i,0)}\wh{S}^{(0,j)}.\label{eq:Sij-dyna-4}
\end{align}
\label{Sij-1-dyna}
\end{subequations}
In addition to  $S^{(i,j)}$,  we define
\begin{equation}
S(a,b)=\bc^T (b\bI+\bK)^{-1}(\bI+\bM)^{-1}(a\bI+\bK)^{-1}\br,~~ a,b\in \mathbb{C},
\label{Sab-1}
\end{equation}
which also has the symmetric property
\begin{equation}
S(a,b)=S(b,a)
\label{Sab-sym}
\end{equation}
and obeys the shift relations
\begin{subequations}
\label{Sab-1-re}
\begin{align}
& 1-(p+b)\wt{S}(a,b)+(p-a)S(a,b)=\wt{V}(a)V(b), \label{Sab-1-re-a}\\
& 1-(q+b)\wh{S}(a,b)+(q-a)S(a,b)=\wh{V}(a)V(b), \label{Sab-1-re-b}
\end{align}
\end{subequations}
where
\begin{eqnarray}
V(a)=1-\bc^T(a \bI+ \bK)^{-1}(\bI+\bM)^{-1}\br=1-\bc^T (\bI+\bM)^{-1}(a\bI+ \bK)^{-1}\br.
\label{Va}
\end{eqnarray}

With the help of the symmetric properties \eqref{Sij-1-sym} and \eqref{Sab-sym},
lattice equations can be obtained as closed forms of the recurrence relations \eqref{Sij-1-dyna} and \eqref{Sab-1-re}.
We list them below:
\begin{itemize}
\item{lpKdV equation  $(w=S^{(0,0)})$:
\begin{equation}
 (p+q+w-\widehat{\widetilde{w}})(p-q+\h{w}-\t{w})=p^2-q^2;\label{lpKdV-a}
\end{equation}
}
\item{lpmKdV equation $(v=1-S^{(0,-1)})$:
\begin{equation}
\label{lpmKdV}
  p(v\wh{v}-\wt{v}\wh{\wt{v}})=q(v\wt{v}-\wh{v}\wh{\wt{v}});
\end{equation}
}
\item{lSKdV equation $(z=S^{(-1,-1)}-\frac{n}{p}-\frac{m}{q})$:
\begin{equation}
\label{lSKdV}
\frac{(z-\wt{z})(\wh{z}-\wh{\wt{z}})}{(z-\wh{z})(\wt{z}-\wh{\wt{z}})}=\frac{q^2}{p^2};
\end{equation}
}
\item{NQC equation:
\begin{equation}\label{NQC}
 \frac{1-(p+b)\wh{\wt{S}}(a,b)+(p-a)\wh{S}(a,b)}
{1-(q+b)\wh{\wt{S}}(a,b)+(q-a)\wt{S}(a,b)}
=\frac{1-(q+a)\wh{S}(a,b)+(q-b)S(a,b)}
{1-(p+a)\wt{S}(a,b)+(p-b)S(a,b)}.
\end{equation}
}
\end{itemize}


\end{document}